# Enlightening the blind spot of the Michaelis–Menten rate law: The role of relaxation dynamics in molecular complex formation

Junghun Chae[1,7], Roktaek Lim[1,2,7], Thomas L. P. Martin[2], Cheol-Min Ghim[1,3,4,*], and Pan-Jun Kim[2,4,5,6,*]

[1]Department of Physics, Ulsan National Institute of Science and Technology, Ulsan 44919, Republic of Korea
[2]Department of Biology, Hong Kong Baptist University, Kowloon, Hong Kong
[3]Department of Biomedical Engineering, Ulsan National Institute of Science and Technology, Ulsan 44919, Republic of Korea
[4]Asia Pacific Center for Theoretical Physics, Pohang, Gyeongbuk 37673, Republic of Korea
[5]Center for Quantitative Systems Biology & Institute of Computational and Theoretical Studies, Hong Kong Baptist University, Kowloon, Hong Kong
[6]Abdus Salam International Centre for Theoretical Physics, 34151 Trieste, Italy
[7]J.C. and R.L. contributed equally to this work.
[*]Author to whom correspondence should be addressed:
C.-M.G. (email: cmghim@unist.ac.kr) or P.-J.K. (email: extutor@gmail.com).

## Abstract

The century-long Michaelis–Menten rate law and its modifications in the modeling of biochemical rate processes stand on the assumption that the concentration of the complex of interacting molecules, at each moment, rapidly approaches an equilibrium (quasi-steady state) compared to the pace of molecular concentration changes. Yet, in the case of actively time-varying molecular concentrations with transient or oscillatory dynamics, the deviation of the complex profile from the quasi-steady state becomes relevant. A recent theoretical approach, known as the effective time-delay scheme (ETS), suggests that the delay by the relaxation time of molecular complex formation contributes to the substantial breakdown of the quasi-steady state assumption. Here, we systematically expand this ETS and inquire into the comprehensive roles of relaxation dynamics in complex formation. Through the modeling of rhythmic protein–protein and protein–DNA interactions and the mammalian circadian clock, our analysis reveals the effect of the relaxation dynamics beyond the time delay, which extends to the dampening of changes in the complex concentration with a reduction in the oscillation amplitude compared to the quasi-steady state. Interestingly, the combined effect of the time delay and amplitude reduction shapes both qualitative and quantitative oscillatory patterns such as the emergence and variability of the mammalian circadian rhythms. These findings highlight the downside of the routine assumption of quasi-steady states and enhance the mechanistic understanding of rich time-varying biomolecular processes.

## I. Introduction

Time-scale separation has been a long practice in the modeling of many physical, chemical, and biological systems where the "fast" components undergo instantaneous equilibrium and only the "slow" components govern the relevant dynamics [1–8]. In the case of biochemical processes, it is often assumed that the concentration of the complex of interacting molecules, at each moment, approaches an equilibrium (quasi-steady state) much faster than the molecular concentrations change [5,7–10]. The Michaelis–Menten (MM) rate law under this quasi-steady state assumption, proposed over a century ago, has been the dominant paradigm of modeling the rates of enzyme-catalyzed reactions [11,12]. The MM rate law has also been widely applied for other molecular interactions in biochemistry and pharmacology, such as reversible protein–protein, transcription factor–gene, and receptor–ligand bindings [13–19].

However, derived by the method called the *standard quasi-steady state approximation* (sQSSA) [5,6,9,20–22], the MM rate law presumes that the enzyme concentration is low enough and thus the concentration of enzyme–substrate complex is negligible compared to substrate concentration. This condition is often satisfied in the case of metabolic reactions, with substrate concentrations typically far higher than the enzyme concentrations. Nevertheless, in the case of protein–protein interactions, the interacting proteins as "enzymes" and "substrates" are often comparable with each other in their concentrations [23–25]. Therefore, the application of the MM rate law for protein–protein interactions has been challenged, with its remedied form from the *total quasi-steady state approximation* (tQSSA) [7,8,10,20,22,26–31]. The tQSSA gives a more accurate description of molecular interactions for a broad range of paired concentrations than the sQSSA, and thus works for protein–protein interactions as well [10,20,26–31]. Later, we will overview the tQSSA and its relationship with the MM rate law from the sQSSA.

Despite the remedy of the MM rate law by the tQSSA, both the tQSSA and sQSSA still rely on the quasi-steady state assumption [6,10,20]. The validity of this assumption across various biochemical systems remains uncertain. Many cellular activities, including circadian or diurnal rhythms, cell cycles, and signal responses [14,15,31–34], exhibit actively time-varying molecular concentrations and therefore may not strictly follow the quasi-steady state assumption.

Recently, we have introduced the mathematical approach that better describes the interaction of time-varying molecular components than the tQSSA and sQSSA [35]. This new approach, termed the effective time-delay scheme (ETS), incorporates into the tQSSA a time-delay effect from the relaxation dynamics of molecular complex formation. With valuable analytical insights, the ETS accounts for simulated and empirical patterns of transient or oscillatory biochemical processes that the quasi-steady state assumption fails to capture [35]. Yet, beyond the time-delay effect, we are prompted to ask whether other aspects of relaxation dynamics also crucially affect the trajectory of the complex concentration. The quest for the role of relaxation dynamics in molecular complex formation will enlighten the blind spot of the quasi-steady state assumption and enhance the mechanistic understanding of rich time-varying biomolecular activities.

To systematically address this issue, we here develop a theoretical framework for the core effects of relaxation dynamics on the complex formation and demonstrate its utility through the analysis of rhythmic protein–protein and protein–DNA interactions and the mammalian circadian system. Our work reveals that the relaxation dynamics does not only delay but also dampens the changes in complex concentration over time by reducing its oscillation amplitude compared to the quasi-steady state. This joint effect of the time delay and amplitude reduction act critical to both the qualitative and quantitative aspects of the molecular oscillatory patterns such as the emergence and variability of the mammalian circadian rhythms.

## II. Theory Overview and Development

Sec. II.A provides an overview of the tQSSA and sQSSA, and the expansion of the ETS to examine the key roles of relaxation dynamics in complex formation. For the ETS, we linearize the differential equation of complex formation and estimate the complex concentration using Eq. (14), the time integral of the tQSSA with an exponential kernel-like function. From Eq. (14), we derive three versions of the ETS—Eqs. (18), (19), and (21). As will be demonstrated later, these versions collectively help dissect different roles of the relaxation dynamics, overlooked by the quasi-steady state assumption.

Sec. II.B focuses on transcription factor–DNA interactions, where inherent stochasticity comes into effect. Starting with the chemical master equation, we derive three versions of the ETS for this case—Eqs. (31)–(33), analogous to those in Sec. II.A.

### A. General formulation

At the beginning, we overview the tQSSA and its relationship with the sQSSA. Consider the scenario where two different molecules A and B bind to each other and form complex AB. For example, the molecules A and B can be two different proteins in heterodimer formation, substrate and enzyme in a metabolic reaction, or solute and transporter in membrane transport. $A(t)$, $B(t)$, and $C(t)$ denote the total concentrations of A and B and the concentration of the complex AB at time $t$, respectively. $A(t)$ and $B(t)$ are not necessarily constant but very generic in their time-dependency, e.g., even with the feedback effect as demonstrated in Sec. III.C later. $C(t)$ is determined by the following equation based on the mass-action law:

$$\frac{\mathrm{d}C(t)}{\mathrm{d}t} = k_{\mathrm{a}}[A(t) - C(t)][B(t) - C(t)] - k_{\delta}C(t), \tag{1}$$

where $A(t) - C(t)$ and $B(t) - C(t)$ indicate the concentrations of free A and B, respectively, $k_{\mathrm{a}}$ is the association rate of free A and B, and $k_{\delta}$ is the effective "decay" rate of AB. For the sake of generality, $k_{\delta}$ is not limited to dissociation but covers all rate processes to lower the level of AB, with $k_{\delta} \equiv k_{\mathrm{d}} + r_{\mathrm{c}} + k_{\mathrm{loc}} + k_{\mathrm{dlt}}$ where $k_{\mathrm{d}}$, $k_{\mathrm{loc}}$, and $k_{\mathrm{dlt}}$ stand for the dissociation, translocation, and dilution rates of AB, respectively, and $r_{\mathrm{c}}$ for the chemical conversion or translocation rate of A or B upon the formation of AB. In this way, $A(t)$, $B(t)$, and $C(t)$ are able to serve any open or closed system.

With notations $\tau \equiv k_{\delta}t$, $K \equiv k_{\delta}/k_{\mathrm{a}}$, $\bar{A}(\tau) \equiv A(t)/K$, $\bar{B}(\tau) \equiv B(t)/K$, and $\bar{C}(\tau) \equiv C(t)/K$, one can rewrite Eq. (1) as

$$\frac{\mathrm{d}\bar{C}(\tau)}{\mathrm{d}\tau} = [\bar{A}(\tau) - \bar{C}(\tau)][\bar{B}(\tau) - \bar{C}(\tau)] - \bar{C}(\tau). \tag{2}$$

By definition, $\overline{C}(\tau) \leq \overline{A}(\tau)$ and $\overline{C}(\tau) \leq \overline{B}(\tau)$, and therefore

$$\overline{C}(\tau) \leq \min\left[\overline{A}(\tau), \overline{B}(\tau)\right]. \tag{3}$$

On the other hand, Eq. (2) is equivalent to

$$\frac{\mathrm{d}\bar{C}(\tau)}{\mathrm{d}\tau} = \left[\bar{C}(\tau) - \bar{C}_{\mathrm{tQ}}(\tau)\right]\left\{\bar{C}(\tau) - \left[\bar{C}_{\mathrm{tQ}}(\tau) + \Delta_{\mathrm{tQ}}(\tau)\right]\right\}, \tag{4}$$

where $\bar{C}_{\mathrm{tQ}}(\tau)$ and $\Delta_{\mathrm{tQ}}(\tau)$ are given by

$$\bar{C}_{\mathrm{tQ}}(\tau) \equiv \frac{1}{2}\left[1 + \overline{A}(\tau) + \overline{B}(\tau) - \Delta_{\mathrm{tQ}}(\tau)\right], \tag{5}$$

$$\begin{aligned} \Delta_{\mathrm{tQ}}(\tau) &\equiv \sqrt{[1 + \overline{A}(\tau) + \overline{B}(\tau)]^2 - 4\overline{A}(\tau)\overline{B}(\tau)} \\ &= \sqrt{1 + 2\left[\overline{A}(\tau) + \overline{B}(\tau)\right] + \left[\overline{A}(\tau) - \overline{B}(\tau)\right]^2}. \end{aligned} \tag{6}$$

Given $\bar{A}(\tau)$ and $\bar{B}(\tau)$, the tQSSA assumes that $\bar{C}(\tau)$ approaches the quasi-steady state fast enough each time [10,20]. This idea can be understood by noticing that $\bar{C}'(\tau) \to 0$ in Eq. (4) when $\bar{C}(\tau) \to \bar{C}_{\mathrm{tQ}}(\tau)$. In addition, $\bar{C}_{\mathrm{tQ}}(\tau) \leq \min\left[\overline{A}(\tau), \overline{B}(\tau)\right]$ and thus Eq. (3) is naturally satisfied when $\bar{C}(\tau) = \bar{C}_{\text{tQ}}(\tau)$. The other nominal solution of $\bar{C}'(\tau) = 0$ in Eq. (4) is physically senseless because it does not satisfy Eq. (3).

In the tQSSA, one takes $\bar{C}_{\text{tQ}}(\tau)$ as the estimate of $\bar{C}(\tau)$. The tQSSA works generally more accurately than the conventional MM rate law [7,8,10,20,26–31]. To obtain the MM rate law, consider the following situation:

$$\overline{B}(\tau) \ll 1 + \overline{A}(\tau) \text{ or } \overline{A}(\tau) \ll 1 + \overline{B}(\tau). \tag{7}$$

In this situation, the Padé approximant for $\bar{C}_{\text{tQ}}(\tau)$ takes the following form:

$$\bar{C}_{\text{tQ}}(\tau) \approx \frac{\overline{A}(\tau)\overline{B}(\tau)}{1+\overline{A}(\tau)+\overline{B}(\tau)}. \tag{8}$$

Eq. (8) renders Eq. (7) similar to condition $\bar{C}_{\text{tQ}}(\tau)/\overline{A}(\tau) \ll 1$ or $\bar{C}_{\text{tQ}}(\tau)/\overline{B}(\tau) \ll 1$. In other words, Eq. (8) would be valid when AB complex concentration is negligible compared to either A or B concentration. Essentially, this condition is identical to the assumption in the sQSSA that gives the MM rate law [5,6]. A typical metabolic reaction with $\overline{B}(\tau) \ll \overline{A}(\tau)$ for substrate A and enzyme B automatically satisfies Eq. (7) and then Eq. (8) reduces to $\bar{C}_{\text{tQ}}(\tau) \approx \overline{A}(\tau)\overline{B}(\tau)/\left[1 + \overline{A}(\tau)\right]$, equivalent to the familiar MM rate law $A(t)B(t)/[K + A(t)]$ for AB concentration [5,6,9,11,12,20,35]. Clearly, $K$ here is the Michaelis constant, commonly known as $K_{\text{M}}$. To be precise, the sQSSA uses the concentration of free A instead of that of the total A, but we refer to the latter case as the sQSSA because of the negligible complex level in the assumption.

Unlike the MM rate law limited to the condition in Eq. (7), the tQSSA is free from that condition and thus more widely applicable [10,20,26–31]. Still, both the tQSSA and sQSSA rely on the assumption that $\overline{C}(\tau)$ fast approaches the quasi-steady state each time before the marked change of $\overline{A}(\tau)$ or $\overline{B}(\tau)$. Next, we relax this quasi-steady state assumption and extend the analysis to the case of time-varying $\overline{A}(\tau)$ and $\overline{B}(\tau)$ [35].

Suppose that $\overline{C}(\tau)$ does not necessarily approach the quasi-steady state, but stays within some distance from it as

$$\left|\bar{C}(\tau) - \bar{C}_{\text{tQ}}(\tau)\right| \ll \Delta_{\text{tQ}}(\tau). \tag{9}$$

We previously showed that this relation is readily satisfied in physiologically-relevant conditions (Table S1) [35]. It allows us to discard $\left[\bar{C}(\tau) - \bar{C}_{\text{tQ}}(\tau)\right]^2$ compared with $\Delta_{\text{tQ}}(\tau)\left[\bar{C}(\tau) - \bar{C}_{\text{tQ}}(\tau)\right]$ and then reduce Eq. (4) to

$$\frac{\mathrm{d}\bar{C}(\tau)}{\mathrm{d}\tau} \approx \Delta_{\text{tQ}}(\tau)\left[\bar{C}_{\text{tQ}}(\tau) - \bar{C}(\tau)\right]. \tag{10}$$

Multiplying both the left- and right-hand sides by $\exp\left[\int_{\tau_0}^{\tau}\Delta_{\mathrm{tQ}}(\tau')\,\mathrm{d}\tau'\right]$ with $\tau_0$ as the initial point of $\tau$ leads to

$$\frac{\mathrm{d}}{\mathrm{d}\tau}\left[\bar{C}(\tau)e^{\int_{\tau_0}^{\tau}\Delta_{\mathrm{tQ}}(\tau')\mathrm{d}\tau'}\right] \approx \Delta_{\mathrm{tQ}}(\tau)\bar{C}_{\mathrm{tQ}}(\tau)e^{\int_{\tau_0}^{\tau}\Delta_{\mathrm{tQ}}(\tau')\mathrm{d}\tau'}. \tag{11}$$

The integral of Eq. (11) from $\tau = \tau_0$ leads to

$$\bar{C}(\tau) \approx \int_{\tau_0}^{\tau}\Delta_{\mathrm{tQ}}(\tau')\bar{C}_{\mathrm{tQ}}(\tau')e^{-\int_{\tau'}^{\tau}\Delta_{\mathrm{tQ}}(\tau'')\mathrm{d}\tau''}\mathrm{d}\tau' + \bar{C}(\tau_0)e^{-\int_{\tau_0}^{\tau}\Delta_{\mathrm{tQ}}(\tau')\mathrm{d}\tau'}. \tag{12}$$

This expression covers both the asymptotic state and the initial condition-dependent transient. Noteworthy is that, in enzyme kinetics, singular perturbation methods have been used for the matching of short- and long-term dynamics as a systematic basis of the sQSSA and tQSSA [5–8].

We now assume that $\Delta_{\mathrm{tQ}}(\tau')$ changes rather slowly over $\tau'$ to satisfy

$$\Delta_{\mathrm{tQ}}(\tau') \approx \Delta_{\mathrm{tQ}}(\tau) \text{ for } \tau' \text{ in the range } \tau - \Delta_{\mathrm{tQ}}^{-1}(\tau) \lesssim \tau' \leq \tau. \tag{13}$$

In physiologically-relevant conditions (Table S1), Eq. (13) is readily satisfied as we previously showed [35]. From Eq. (13), $\int_{\tau'}^{\tau}\Delta_{\mathrm{tQ}}(\tau'')\mathrm{d}\tau'' \approx (\tau-\tau')\Delta_{\mathrm{tQ}}(\tau)$ and $\int_{\tau_0}^{\tau}\Delta_{\mathrm{tQ}}(\tau')\,\mathrm{d}\tau' \approx (\tau-\tau_0)\Delta_{\mathrm{tQ}}(\tau_0)$ for $\tau' \gtrsim \tau - \Delta_{\mathrm{tQ}}^{-1}(\tau)$ and $\tau \lesssim \tau_0 + \Delta_{\mathrm{tQ}}^{-1}(\tau_0)$, respectively. Hence, Eq. (12) for $\tau \gg \tau_0 + \Delta_{\mathrm{tQ}}^{-1}(\tau_0)$ reaches the following form:

$$\bar{C}(\tau) \approx \Delta_{\mathrm{tQ}}(\tau)\int_{-\infty}^{\tau}\bar{C}_{\mathrm{tQ}}(\tau')e^{-(\tau-\tau')\Delta_{\mathrm{tQ}}(\tau)}\mathrm{d}\tau'. \tag{14}$$

The right-hand side is not sensitive to the lower bound of $\tau'$ for the integral as long as this lower bound is $\ll \tau - \Delta_{\mathrm{tQ}}^{-1}(\tau)$.

In Eq. (14), the effect of $\bar{C}_{\mathrm{tQ}}(\tau')$ exponentially decays over time by $\exp\left[-(\tau-\tau')\Delta_{\mathrm{tQ}}(\tau)\right]$ and its decay time-scale $\Delta_{\mathrm{tQ}}^{-1}(\tau)$ approximates molecular relaxation time during which the effect of instantaneous A and B levels through $\bar{C}_{\mathrm{tQ}}(\tau')$ is notably sustained in AB complex formation. Intuitively from Eq. (14), the quasi-steady state assumption $\bar{C}(\tau) \approx \bar{C}_{\mathrm{tQ}}(\tau)$ is valid when $\bar{C}_{\mathrm{tQ}}(\tau') \approx \bar{C}_{\mathrm{tQ}}(\tau)$ [i.e., $\left|\bar{C}_{\mathrm{tQ}}(\tau') - \bar{C}_{\mathrm{tQ}}(\tau)\right| \ll \bar{C}_{\mathrm{tQ}}(\tau)$] for $\tau - \tau' \lesssim \Delta_{\mathrm{tQ}}^{-1}(\tau)$. In other words,

$$1 \gg \frac{\Delta_{\mathrm{tQ}}^{-1}(\tau)}{\bar{C}_{\mathrm{tQ}}(\tau)}\left|\frac{\mathrm{d}\bar{C}_{\mathrm{tQ}}(\tau)}{\mathrm{d}\tau}\right| = \frac{\Delta_{\mathrm{tQ}}^{-2}(\tau)}{\bar{C}_{\mathrm{tQ}}(\tau)}\left|\left[\bar{B}(\tau) - \bar{C}_{\mathrm{tQ}}(\tau)\right]\frac{\mathrm{d}\bar{A}(\tau)}{\mathrm{d}\tau} + \left[\bar{A}(\tau) - \bar{C}_{\mathrm{tQ}}(\tau)\right]\frac{\mathrm{d}\bar{B}(\tau)}{\mathrm{d}\tau}\right|, \tag{15}$$

where the equality comes from Eq. (5). This condition means that the relaxation time should be much shorter than the time-scale with which the complex concentration changes. The more precise quasi-steady condition considering up to $\bar{C}''_{\mathrm{tQ}}(\tau)$ was derived in our previous study [35].

To cover more general cases than the quasi-steady condition, we return to Eq. (14) and estimate $\bar{C}(\tau)$ as

$$\bar{C}(\tau) \approx \int_0^1 \left\{\bar{C}_{\mathrm{tQ}}(\tau) \int_0^x e^{-\xi} \mathrm{d}\xi + \bar{C}_{\mathrm{tQ}}\left[\tau - \Delta_{\mathrm{tQ}}^{-1}(\tau)\right] \int_x^\infty e^{-\xi} \mathrm{d}\xi\right\} p(x) \mathrm{d}x$$
$$= \int_0^1 \left\{(1 - e^{-x})\bar{C}_{\mathrm{tQ}}(\tau) + e^{-x}\bar{C}_{\mathrm{tQ}}\left[\tau - \Delta_{\mathrm{tQ}}^{-1}(\tau)\right]\right\} p(x) \mathrm{d}x. \quad (16)$$

Here, $\xi \equiv (\tau - \tau')\Delta_{\mathrm{tQ}}(\tau)$ and $x$ is chosen from $0 \le x < 1$ to roughly satisfy $\bar{C}_{\mathrm{tQ}}(\tau') \approx \bar{C}_{\mathrm{tQ}}(\tau)$ for $\tau'$ in the range $\tau - x\Delta_{\mathrm{tQ}}^{-1}(\tau) \le \tau' \le \tau$ and $\bar{C}_{\mathrm{tQ}}(\tau') \approx \bar{C}_{\mathrm{tQ}}\left[\tau - \Delta_{\mathrm{tQ}}^{-1}(\tau)\right]$ for $\tau'$ in the range $\tau - \Delta_{\mathrm{tQ}}^{-1}(\tau) \lesssim \tau' < \tau - x\Delta_{\mathrm{tQ}}^{-1}(\tau)$. $x$ depends on the specific form of $\bar{C}_{\mathrm{tQ}}(\tau')$, and thus $p(x)$ is introduced above as the likelihood of particular $x$ across all probable forms of $\bar{C}_{\mathrm{tQ}}(\tau')$ with condition $\int_0^1 p(x)\mathrm{d}x = 1$. Given an uncertain distribution of the form of $\bar{C}_{\mathrm{tQ}}(\tau')$, it would be safe to assume near-uniformity of $p(x)$ with $p(x) \simeq 1$ for $0 \le x < 1$, i.e., equal *a priori* probabilities. Therefore, $\bar{C}(\tau)$ in Eq. (16) is further estimated as

$$\bar{C}(\tau) \approx e^{-1}\bar{C}_{\mathrm{tQ}}(\tau) + (1 - e^{-1})\bar{C}_{\mathrm{tQ}}\left[\tau - \Delta_{\mathrm{tQ}}^{-1}(\tau)\right]. \quad (17)$$

Obviously here, the molecular relaxation time $\Delta_{\mathrm{tQ}}^{-1}(\tau)$ contributes a time delay to complex formation. At a completely steady state, $\bar{C}_{\mathrm{tQ}}(\tau) = \bar{C}_{\mathrm{tQ}}\left[\tau - \Delta_{\mathrm{tQ}}^{-1}(\tau)\right]$ and thus $\bar{C}(\tau)$ becomes the same as $\bar{C}_{\mathrm{tQ}}(\tau)$, i.e., the tQSSA.

Because $1 - e^{-1}$ is about twice the value of $e^{-1}$, one may further simplify Eq. (17) to $\bar{C}(\tau) \approx \bar{C}_{\mathrm{tQ}}\left[\tau - \Delta_{\mathrm{tQ}}^{-1}(\tau)\right]$ by leaving the last term alone with its weight increased from $1 - e^{-1}$ to 1 for the compensation. One caveat with this approximation is that $\bar{C}_{\mathrm{tQ}}\left[\tau - \Delta_{\mathrm{tQ}}^{-1}(\tau)\right]$ may not necessarily satisfy the relation $\bar{C}_{\mathrm{tQ}}\left[\tau - \Delta_{\mathrm{tQ}}^{-1}(\tau)\right] \le \min\left[\bar{A}(\tau), \bar{B}(\tau)\right]$ favored by Eq. (3). As a practical safeguard to avoid this problem, we have recently proposed the following approximant for $\bar{C}(\tau)$ consistent with Eq. (3) [35]:

$$\bar{C}_{\gamma_1}(\tau) \equiv \min\left\{\bar{C}_{\mathrm{tQ}}\left[\tau - \Delta_{\mathrm{tQ}}^{-1}(\tau)\right], \bar{A}(\tau), \bar{B}(\tau)\right\}. \quad (18)$$

Although the above $\bar{C}_{\gamma_1}(\tau)$ looks rather complex, this form is essentially a simple conversion $\tau \to \tau - \Delta_{\mathrm{tQ}}^{-1}(\tau)$ in the tQSSA and $\min\{\cdots\}$ is just taken for minor correction. Of note, our previous derivation [35] is based on the Taylor expansion of $\bar{C}_{\mathrm{tQ}}(\tau')$ in Eq. (14) by $\tau - \tau'$, not the above procedure itself.

Alternatively, we propose the following new approximant for $\bar{C}(\tau)$ from the entire Eq. (17) without any reduction:

$$\bar{C}_{\gamma_2}(\tau) \equiv \min\left\{e^{-1}\bar{C}_{\mathrm{tQ}}(\tau) + (1-e^{-1})\bar{C}_{\mathrm{tQ}}\left[\tau - \Delta_{\mathrm{tQ}}^{-1}(\tau)\right], \bar{A}(\tau), \bar{B}(\tau)\right\}. \tag{19}$$

Both $\bar{C}_{\gamma_1}(\tau)$ and $\bar{C}_{\gamma_2}(\tau)$ in Eqs. (18) and (19) include an explicit time delay $\Delta_{\mathrm{tQ}}^{-1}(\tau)$. We refer to this type of the formulation as the effective time-delay scheme (ETS) as in our previous study [35]. The ETS is the generalization of the MM rate law for time-varying molecular concentrations that may not strictly adhere to the quasi-steady state assumption. More specifically, we will refer to the use of Eq. (18) as the $\mathrm{ETS}_1$ and that of Eq. (19) as the $\mathrm{ETS}_2$. The $\mathrm{ETS}_2$ is newly introduced here, and tends to be more accurate than the $\mathrm{ETS}_1$ as we will elaborate later. If the relaxation time in the complex formation is so short that the effective time delay in Eqs. (18) and (19) can be ignored, both the $\mathrm{ETS}_1$ and $\mathrm{ETS}_2$ return to the tQSSA in their forms.

About the physical interpretation of the time delay $\Delta_{\mathrm{tQ}}^{-1}(\tau)$, we previously proved that it is inversely related to free molecule availability [35], as $\Delta_{\mathrm{tQ}}^{-1}(\tau) = \left\{1 + \overline{A}(\tau) + \overline{B}(\tau) - 2\bar{C}_{\mathrm{tQ}}(\tau)\right\}^{-1}$ from Eq. (5). Here, $\overline{A}(\tau) + \overline{B}(\tau) - 2\bar{C}_{\mathrm{tQ}}(\tau) = \left[\overline{A}(\tau) - \bar{C}_{\mathrm{tQ}}(\tau)\right] + \left[\overline{B}(\tau) - \bar{C}_{\mathrm{tQ}}(\tau)\right]$, which is proportional to the total free molecule concentration at the quasi-steady state. In other words, the less the free molecules, the more the time delay. One can understand this observation as follows: over the course of complex formation, free A and B are getting depleted and therefore the rate of complex formation $[\bar{A}(\tau) - \bar{C}(\tau)][\bar{B}(\tau) - \bar{C}(\tau)]$ in Eq. (2) continues to decline towards quicker relaxation of the complex concentration. This free-molecule depletion effect to shorten the relaxation time is proportional to the free molecule concentration itself because Eq. (10) is rewritten as $\left[\bar{C}(\tau) - \bar{C}_{\mathrm{tQ}}(\tau)\right]' \approx -\bar{C}'_{\mathrm{tQ}}(\tau) - \left[\Delta_{\mathrm{tQ}}(\tau) - 1\right]\left[\bar{C}(\tau) - \bar{C}_{\mathrm{tQ}}(\tau)\right] - \left[\bar{C}(\tau) - \bar{C}_{\mathrm{tQ}}(\tau)\right]$ and the second term on the right-hand side here is proportional to $\Delta_{\mathrm{tQ}}(\tau) - 1$. Therefore, the relaxation time takes a decreasing function of the free molecule concentration, consistent with the above observation. Clearly, the free molecule concentration would be low for relatively few A and B molecules with comparable concentrations—i.e., small $\overline{A}(\tau) + \overline{B}(\tau)$ and $[\overline{A}(\tau) - \overline{B}(\tau)]^2$ in Eq. (6). In this case, the relaxation time would be relatively long and the tQSSA or sQSSA cannot correctly describe that situation. For example, protein–protein interactions would often be the cases that require the consideration of the effective time delay, compared to metabolic reactions with much excess substrates not binding to enzymes.

In the case of oscillating molecular concentrations, it is straightforward to expect that the $\mathrm{ETS}_1$ and tQSSA would exhibit temporal profiles with almost the same amplitudes but a certain phase difference between them. The $\mathrm{ETS}_2$, which is likely more correct than the $\mathrm{ETS}_1$ in its derivation, would also give the phase difference but relatively small, due to the superposition of both the time-delayed and -undelayed forms of the tQSSA. Furthermore, in contrast to the $\mathrm{ETS_1}$, the amplitude from the $\mathrm{ETS_2}$ is reduced by this superposition and thus smaller than the tQSSA's, as proven in Appendix A. These results suggest that the relaxation dynamics does not only delay but also dampens the changes in complex concentration over time, compared to its quasi-steady state.

In the case of periodic molecular oscillations, it is possible to gain a more systematic insight into the above delay and amplitude reduction. Applying the inverse Fourier transform $\bar{C}_{\mathrm{tQ}}(\tau) = \int_{-\infty}^{\infty} \bar{C}_{\mathrm{tQ}}^{\mathrm{f}}(\xi) e^{2\pi i \tau \xi} \mathrm{d}\xi$ to Eq. (14) results in

$$\begin{aligned}\bar{C}(\tau) &\approx \Delta_{\mathrm{tQ}}(\tau) e^{-\Delta_{\mathrm{tQ}}(\tau)\tau} \int_{-\infty}^{\infty} \int_{-\infty}^{\tau} \bar{C}_{\mathrm{tQ}}^{\mathrm{f}}(\xi) e^{[2\pi i \xi + \Delta_{\mathrm{tQ}}(\tau)]\tau'} \mathrm{d}\tau' \mathrm{d}\xi \\ &= \int_{-\infty}^{\infty} \frac{\bar{C}_{\mathrm{tQ}}^{\mathrm{f}}(\xi)}{\sqrt{1+\left[2\pi\xi\Delta_{\mathrm{tQ}}^{-1}(\tau)\right]^2}} e^{2\pi i \xi\left\{\tau - \frac{1}{2\pi\xi}\arctan\left[2\pi\xi\Delta_{\mathrm{tQ}}^{-1}(\tau)\right]\right\}} \mathrm{d}\xi. \end{aligned} \tag{20}$$

Because $\bar{C}_{\mathrm{tQ}}^{\mathrm{f}}(\xi)$ is the Fourier transform of $\bar{C}_{\mathrm{tQ}}(\tau)$, the above formula resembles the $\mathrm{ETS_1}$ but with a modified time delay $\arctan\left[2\pi\xi\Delta_{\mathrm{tQ}}^{-1}(\tau)\right]/2\pi\xi$ and the amplitude reduced by a factor of $\sqrt{1+\left[2\pi\xi\Delta_{\mathrm{tQ}}^{-1}(\tau)\right]^2}$. This observation suggests the following approximant for $\bar{C}(\tau)$ when the molecular concentrations oscillate with a constant period $T$:

$$\bar{C}_{\gamma_3}(\tau) \equiv \min\left\{ \frac{\bar{C}_{\mathrm{tQ}}\left\{\tau - \frac{k_\delta T}{2\pi}\arctan\left[\frac{2\pi\Delta_{\mathrm{tQ}}^{-1}(\tau)}{k_\delta T}\right]\right\} - \langle \bar{C}_{\mathrm{tQ}}(\tau)\rangle}{\sqrt{1+\left[\frac{2\pi\Delta_{\mathrm{tQ}}^{-1}(\tau)}{k_\delta T}\right]^2}} + \langle \bar{C}_{\mathrm{tQ}}(\tau)\rangle, \bar{A}(\tau), \bar{B}(\tau) \right\}, \tag{21}$$

where $\langle \bar{C}_{\mathrm{tQ}}(\tau)\rangle$ is the time average of $\bar{C}_{\mathrm{tQ}}(\tau)$ over the oscillation period. The delay $(k_\delta T/2\pi)\cdot\arctan\left[2\pi\Delta_{\mathrm{tQ}}^{-1}(\tau)/k_\delta T\right]$ coincides with $\Delta_{\mathrm{tQ}}^{-1}(\tau)$ at small $2\pi\Delta_{\mathrm{tQ}}^{-1}(\tau)/k_\delta T$ and becomes saturated towards $k_\delta T/4$. We will refer to the use of Eq. (21) as the $\mathrm{ETS_3}$. Unlike the $\mathrm{ETS_1}$ and $\mathrm{ETS_2}$, the $\mathrm{ETS_3}$ is only applicable to a periodic oscillation of complex concentration. Yet, in that case, the $\mathrm{ETS_3}$ predicts the oscillation phase most accurately with its delay term as we will see later, and also reveals the correct analytical dependency of the amplitude reduction on the relaxation time.

Table S2 provides a comparative overview of the $\mathrm{ETS_1}$, $\mathrm{ETS_2}$, and $\mathrm{ETS_3}$. As we will see later, these three can be used together to dissect different roles of relaxation dynamics in complex formation, which are not captured by the quasi-steady state assumption. For example, the comparison of the $\mathrm{ETS_1}$ to the tQSSA can reveal the effect of the time delay from the relaxation dynamics, while the result of the $\mathrm{ETS_2}$ further reflects the effect of the amplitude reduction. The precise mathematical condition for the validity of the $ETS_1$ is already available [35], and the $ETS_2$ and $ETS_3$ tend to be more correct than the $ETS_1$ according to our analysis later.

**B. Transcription factor–DNA binding case**

Thus far, we have implicitly assumed the continuity of molecular concentrations as in Eq. (1). However, there exist biomolecular events much away from this assumption. For example, a transcription factor (TF) binds to a DNA molecule in the nucleus to regulate mRNA production and the number of this TF–DNA assembly would be either 0 or 1 for a DNA site that can at most afford one copy of the TF. The inherently discrete and stochastic nature of the TF–DNA binding looks contrasted with the continuous and deterministic nature of Eq. (1). To rigorously handle this TF–DNA binding case, we will now use the chemical master equation [36].

$P(n,t)$ denotes the probability that $n$ copies of the TF occupy the DNA site at time $t$. If this DNA site can at most afford $N$ copies of the TF at once, $n = 0, 1, \cdots, N$ and $\sum_{n=0}^{N} P(n,t) = 1$. If we further define $P(n,t) \equiv 0$ for $n \neq 0, 1, \cdots, N$ and assume that the DNA-binding TFs are hardly accessible by proteolytic machineries for their degradation, the dynamics of $P(n,t)$ with $n = 0, 1, \cdots, N$ is governed by the following master equation:

$$\frac{\partial P(n,t)}{\partial t} = k_{\mathrm{TFa}} V \left[A_{\mathrm{TF}}(t) - \frac{n-1}{V}\right]\left(B_{\mathrm{DNA}} - \frac{n-1}{V}\right) P(n-1,t) - \left\{k_{\mathrm{TFa}} V \left[A_{\mathrm{TF}}(t) - \frac{n}{V}\right]\left(B_{\mathrm{DNA}} - \frac{n}{V}\right) + n k_{\mathrm{TF\delta}}\right\} P(n,t) + (n+1) k_{\mathrm{TF\delta}} P(n+1,t), \quad (22)$$

where $k_{\mathrm{TFa}}$ and $k_{\mathrm{TF\delta}}$ denote the TF–DNA binding and unbinding rates, respectively, $V$ is the nuclear volume, $A_{\mathrm{TF}}(t)$ is the total TF concentration in the nucleus, and $B_{\mathrm{DNA}}$ is the "concentration" of the target DNA site, i.e., $B_{\mathrm{DNA}} = NV^{-1}$. Here, we assume that $A_{\mathrm{TF}}(t)$ is uniquely determined at each time $t$ with little stochasticity in $A_{\mathrm{TF}}(t)$ itself and the nuclear volume $V$ is constant.

The ensemble average $C_{\mathrm{TF}}(t) \equiv \langle nV^{-1} \rangle = V^{-1} \sum_{n=0}^{N} nP(n,t)$ is expressed by Eq. (22) as

$$\begin{aligned}\frac{\mathrm{d}C_{\mathrm{TF}}(t)}{\mathrm{d}t} &= k_{\mathrm{TFa}} \sum_{n=0}^{N} \left[A_{\mathrm{TF}}(t) - \frac{n}{V}\right]\left(B_{\mathrm{DNA}} - \frac{n}{V}\right) P(n,t) - k_{\mathrm{TF\delta}} C_{\mathrm{TF}}(t) \\ &= k_{\mathrm{TFa}} \left\{[A_{\mathrm{TF}}(t) - C_{\mathrm{TF}}(t)][B_{\mathrm{DNA}} - C_{\mathrm{TF}}(t)] + \left\langle \left(\frac{n}{V}\right)^2 \right\rangle - \left\langle \frac{n}{V} \right\rangle^2 \right\} - k_{\mathrm{TF\delta}} C_{\mathrm{TF}}(t), \quad (23)\end{aligned}$$

where $\langle (nV^{-1})^2 \rangle = V^{-2} \sum_{n=0}^{N} n^2 P(n,t)$. Eq. (23) resembles Eq. (1), in the absence of the stochastic fluctuation in the TF binding [$\langle (nV^{-1})^2 \rangle - \langle nV^{-1} \rangle^2$]. However, the stochastic fluctuation cannot be discarded for small $N$. For simplicity, we focus on $N = 1$ and thus $B_{\mathrm{DNA}} = V^{-1}$. Eq. (23) is then rewritten as

$$\frac{\mathrm{d}\bar{C}_{\mathrm{TF}}(\tau)}{\mathrm{d}\tau} = \frac{\bar{A}_{\mathrm{TF}}(\tau)}{K_{\mathrm{TF}}V} - [1 + \bar{A}_{\mathrm{TF}}(\tau)]\bar{C}_{\mathrm{TF}}(\tau) = [1 + \bar{A}_{\mathrm{TF}}(\tau)]\left[\bar{C}_{\mathrm{TFQ}}(\tau) - \bar{C}_{\mathrm{TF}}(\tau)\right], \quad (24)$$

where $\tau \equiv k_{\mathrm{TF\delta}}t$, $K_{\mathrm{TF}} \equiv k_{\mathrm{TF\delta}}/k_{\mathrm{TFa}}$, $\bar{C}_{\mathrm{TF}}(\tau) \equiv C_{\mathrm{TF}}(t)/K_{\mathrm{TF}}$, $\bar{A}_{\mathrm{TF}}(\tau) \equiv A_{\mathrm{TF}}(t)/K_{\mathrm{TF}}$, and

$$\bar{C}_{\mathrm{TFQ}}(\tau) \equiv \frac{\bar{A}_{\mathrm{TF}}(\tau)}{K_{\mathrm{TF}}V[1+\bar{A}_{\mathrm{TF}}(\tau)]}. \quad (25)$$

Because $\bar{C}'_{\mathrm{TF}}(\tau) \to 0$ in Eq. (24) when $\bar{C}_{\mathrm{TF}}(\tau) \to \bar{C}_{\mathrm{TFQ}}(\tau)$ given $\bar{A}_{\mathrm{TF}}(\tau)$, we estimate $\bar{C}_{\mathrm{TF}}(\tau)$ as $\bar{C}_{\mathrm{TFQ}}(\tau)$ under the quasi-steady state assumption. In fact, $\bar{C}_{\mathrm{TFQ}}(\tau)$ matches a special case of the preexisting, *stochastic quasi-steady state approximation* (stochastic QSSA) [37,38]. Interestingly, $\bar{C}_{\mathrm{TFQ}}(\tau)$ resembles the MM rate law with the "concentration" of the DNA site $V^{-1}$. $\bar{C}_{\mathrm{TFQ}}(\tau)$, nevertheless, is not a mere continuum of Eq. (8), because the denominator in $\bar{C}_{\mathrm{TFQ}}(\tau)$ has $1 + \bar{A}_{\mathrm{TF}}(\tau)$ but not $1 + \bar{A}_{\mathrm{TF}}(\tau) + (K_{\mathrm{TF}}V)^{-1}$. This discrepancy between $\bar{C}_{\mathrm{TFQ}}(\tau)$ and Eq. (8) roots in the stochasticity of the TF–DNA binding and relates to the previously-reported denominator $1 + \overline{A}(\tau) + \overline{B}(\tau) - (KV)^{-1}$ for the stochastic version of the MM rate law [39]. Together, $\bar{C}_{\mathrm{TFQ}}(\tau)$ is fundamentally more correct for the DNA-binding TF concentration than both the tQSSA and sQSSA in Eqs. (5) and (8), and we will just refer to $\bar{C}_{\mathrm{TFQ}}(\tau)$ as the QSSA for TF–DNA interactions [35].

Still, the use of $\bar{C}_{\mathrm{TFQ}}(\tau)$ relies on the quasi-steady state assumption. We then relax this assumption and extend our analysis to the case of time-varying TF concentration. We start with the exact solution of Eq. (24):

$$\bar{C}_{\mathrm{TF}}(\tau) = \int_{\tau_0}^{\tau}[1 + \bar{A}_{\mathrm{TF}}(\tau')]\bar{C}_{\mathrm{TFQ}}(\tau')e^{-\int_{\tau'}^{\tau}[1+\bar{A}_{\mathrm{TF}}(\tau'')]\mathrm{d}\tau''}\mathrm{d}\tau' + \bar{C}_{\mathrm{TF}}(\tau_0)e^{-\int_{\tau_0}^{\tau}[1+\bar{A}_{\mathrm{TF}}(\tau')]\mathrm{d}\tau'}, \quad (26)$$

where $\tau_0$ is the initial point of $\tau$. Assume that $1 + \bar{A}_{\mathrm{TF}}(\tau')$ changes rather slowly over $\tau'$ to satisfy the following relation:

$$1 + \bar{A}_{\mathrm{TF}}(\tau') \approx 1 + \bar{A}_{\mathrm{TF}}(\tau) \text{ for } \tau' \text{ in the range } \tau - \frac{1}{1+\bar{A}_{\mathrm{TF}}(\tau)} \lesssim \tau' \leq \tau. \quad (27)$$

In physiologically-relevant conditions (Table S1), this relation is readily satisfied as we previously showed [35]. With Eq. (27), Eq. (26) for $\tau \gg \tau_0 + [1 + \bar{A}_{\mathrm{TF}}(\tau_0)]^{-1}$ leads to

$$\bar{C}_{\mathrm{TF}}(\tau) \approx [1 + \bar{A}_{\mathrm{TF}}(\tau)]\int_{-\infty}^{\tau}\bar{C}_{\mathrm{TFQ}}(\tau')e^{-[1+\bar{A}_{\mathrm{TF}}(\tau)](\tau-\tau')}\mathrm{d}\tau'. \quad (28)$$

The right-hand side is not sensitive to the lower bound of $\tau'$ for the integral as long as this lower bound is $\ll \tau - [1 + \bar{A}_{\mathrm{TF}}(\tau)]^{-1}$.

In Eq. (28), the effect of $\bar{C}_{\mathrm{TFQ}}(\tau')$ exponentially decays over time by $\exp\{-[1+\bar{A}_{\mathrm{TF}}(\tau)](\tau-\tau')\}$ and its decay time-scale $[1+\bar{A}_{\mathrm{TF}}(\tau)]^{-1}$ approximates molecular relaxation time during which the effect of an instantaneous TF level through $\bar{C}_{\mathrm{TFQ}}(\tau')$ is notably sustained in the TF–DNA assembly. Intuitively from Eq. (28), the quasi-steady state assumption $\bar{C}_{\mathrm{TF}}(\tau) \approx \bar{C}_{\mathrm{TFQ}}(\tau)$ is valid when $\bar{C}_{\mathrm{TFQ}}(\tau') \approx \bar{C}_{\mathrm{TFQ}}(\tau)$ [i.e., $\left|\bar{C}_{\mathrm{TFQ}}(\tau')-\bar{C}_{\mathrm{TFQ}}(\tau)\right| \ll \bar{C}_{\mathrm{TFQ}}(\tau)$] for $\tau-\tau' \lesssim [1+\bar{A}_{\mathrm{TF}}(\tau)]^{-1}$. In other words,

$$1 \gg \frac{[1+\bar{A}_{\mathrm{TF}}(\tau)]^{-1}}{\bar{C}_{\mathrm{TFQ}}(\tau)}\left|\frac{\mathrm{d}\bar{C}_{\mathrm{TFQ}}(\tau)}{\mathrm{d}\tau}\right| = \frac{1}{[1+\bar{A}_{\mathrm{TF}}(\tau)]^2\bar{A}_{\mathrm{TF}}(\tau)}\left|\frac{\mathrm{d}\bar{A}_{\mathrm{TF}}(\tau)}{\mathrm{d}\tau}\right|, \tag{29}$$

where the equality comes from Eq. (25). This condition means that the relaxation time should be much shorter than the time-scale with which the TF–DNA assembly level changes. The more precise quasi-steady condition considering up to $\bar{C}''_{\mathrm{TFQ}}(\tau)$ was derived in our previous study [35].

To cover more general cases than the quasi-steady condition, we return to Eq. (28) and estimate $\bar{C}_{\mathrm{TF}}(\tau)$ as

$$\begin{aligned}\bar{C}_{\mathrm{TF}}(\tau) &\approx \int_0^1 \left\{\bar{C}_{\mathrm{TFQ}}(\tau)\int_0^x e^{-\xi}\mathrm{d}\xi + \bar{C}_{\mathrm{TFQ}}\left[\tau-\frac{1}{1+\bar{A}_{\mathrm{TF}}(\tau)}\right]\int_x^\infty e^{-\xi}\mathrm{d}\xi\right\}p(x)\mathrm{d}x \\ &= \int_0^1 \left\{(1-e^{-x})\bar{C}_{\mathrm{TFQ}}(\tau) + e^{-x}\bar{C}_{\mathrm{TFQ}}\left[\tau-\frac{1}{1+\bar{A}_{\mathrm{TF}}(\tau)}\right]\right\}p(x)\mathrm{d}x. \end{aligned} \tag{30}$$

Here, $\xi \equiv [1+\bar{A}_{\mathrm{TF}}(\tau)](\tau-\tau')$ and $x$ is chosen from $0 \le x < 1$ to roughly satisfy $\bar{C}_{\mathrm{TFQ}}(\tau') \approx \bar{C}_{\mathrm{TFQ}}(\tau)$ for $\tau'$ in the range $\tau - x[1+\bar{A}_{\mathrm{TF}}(\tau)]^{-1} \le \tau' \le \tau$ and $\bar{C}_{\mathrm{TFQ}}(\tau') \approx \bar{C}_{\mathrm{TFQ}}\{\tau-[1+\bar{A}_{\mathrm{TF}}(\tau)]^{-1}\}$ for $\tau'$ in the range $\tau-[1+\bar{A}_{\mathrm{TF}}(\tau)]^{-1} \lesssim \tau' < \tau - x[1+\bar{A}_{\mathrm{TF}}(\tau)]^{-1}$. $x$ depends on the specific form of $\bar{C}_{\mathrm{TFQ}}(\tau')$, and thus $p(x)$ is introduced above as the likelihood of particular $x$ across all probable forms of $\bar{C}_{\mathrm{TFQ}}(\tau')$ with condition $\int_0^1 p(x)\mathrm{d}x = 1$. Given an uncertain distribution of the form of $\bar{C}_{\mathrm{TFQ}}(\tau')$, it would be safe to assume near-uniformity of $p(x)$ with $p(x) \simeq 1$ for $0 \le x < 1$, i.e., equal *a priori* probabilities. In a similar fashion to Sec. II.A, we therefore obtain the following two approximants for $\bar{C}_{\mathrm{TF}}(\tau)$:

$$\bar{C}_{\mathrm{TF}\gamma_1}(\tau) \equiv \bar{C}_{\mathrm{TFQ}}\left[\tau-\frac{1}{1+\bar{A}_{\mathrm{TF}}(\tau)}\right], \tag{31}$$

$$\bar{C}_{\mathrm{TF}\gamma_2}(\tau) \equiv e^{-1}\bar{C}_{\mathrm{TFQ}}(\tau) + (1-e^{-1})\bar{C}_{\mathrm{TFQ}}\left[\tau-\frac{1}{1+\bar{A}_{\mathrm{TF}}(\tau)}\right]. \tag{32}$$

$\bar{C}_{\mathrm{TF}\gamma_1}(\tau)$ and $\bar{C}_{\mathrm{TF}\gamma_2}(\tau)$ represent the TF–DNA versions of the $\mathrm{ETS}_1$ and $\mathrm{ETS}_2$, respectively. Their time-delay term has a similar physical interpretation to that in Sec. II.A. Additionally, this term equals the probability of the DNA unoccupancy at the quasi-steady state, based on Eq. (25). Of note, the $\text{ETS}_1$ with $\bar{C}_{\text{TF}\gamma_1}(\tau)$ was derived in our previous study [35] whereas the $\text{ETS}_2$ with $\bar{C}_{\text{TF}\gamma_2}(\tau)$ is newly proposed here. The $\text{ETS}_2$ tends to be more correct than the $\text{ETS}_1$, as we will elaborate later.

In the case of oscillating TF concentration, the $\text{ETS}_1$ and QSSA would have the TF–DNA assembly profiles with the same amplitude but a certain phase difference between them. The $\text{ETS}_2$ would also give the phase difference but relatively small, due to the superposition of both the time-delayed and -undelayed forms of the QSSA. Furthermore, unlike the case of the $\text{ETS}_1$, the amplitude from the $\text{ETS}_2$ is smaller than the QSSA's, as proven in Appendix A. These results suggest that the relaxation dynamics does not only delay but also dampens the changes in the TF–DNA assembly level over time, compared to its quasi-steady state.

In the case of periodic oscillations, one can obtain a more systematic insight into the above delay and amplitude reduction. Like Sec. II.A, the inverse Fourier transform of $\bar{C}_{\text{TFQ}}(\tau)$ in Eq. (28) leads to the following approximant for $\bar{C}_{\text{TF}}(\tau)$ when the TF concentration oscillates with a constant period $T$:

$$\bar{C}_{\text{TF}\gamma_3}(\tau) \equiv \frac{\bar{C}_{\text{TFQ}}\left\{\tau - \frac{k_{\text{TF}\delta}T}{2\pi}\arctan\left\{\frac{2\pi}{k_{\text{TF}\delta}T}[1+\bar{A}_{\text{TF}}(\tau)]^{-1}\right\}\right\} - \langle \bar{C}_{\text{TFQ}}(\tau)\rangle}{\sqrt{1+\left\{\frac{2\pi}{k_{\text{TF}\delta}T}[1+\bar{A}_{\text{TF}}(\tau)]^{-1}\right\}^2}} + \langle \bar{C}_{\text{TFQ}}(\tau)\rangle, \qquad (33)$$

where $\langle \bar{C}_{\text{TFQ}}(\tau)\rangle$ is the time average of $\bar{C}_{\text{TFQ}}(\tau)$ over the oscillation period. The delay $(k_{\text{TF}\delta}T/2\pi)\cdot\arctan\{2\pi[1+\bar{A}_{\text{TF}}(\tau)]^{-1}/k_{\text{TF}\delta}T\}$ coincides with $[1+\bar{A}_{\text{TF}}(\tau)]^{-1}$ at small $2\pi[1+\bar{A}_{\text{TF}}(\tau)]^{-1}/k_{\text{TF}\delta}T$ and becomes saturated towards $k_{\text{TF}\delta}T/4$. $\bar{C}_{\text{TF}\gamma_3}(\tau)$ represents the TF–DNA version of the $\text{ETS}_3$ introduced in Sec. II.A. Unlike the $\text{ETS}_1$ and $\text{ETS}_2$, the $\text{ETS}_3$ is only applicable to a periodic TF oscillation. Yet, in this case, the $\text{ETS}_3$ predicts the oscillation phase most accurately with its delay term as we will see later, and also reveals the correct analytical dependency of the amplitude reduction on the relaxation time.

Table S2 provides a comparative overview of the $\text{ETS}_1$, $\text{ETS}_2$, and $\text{ETS}_3$. As we will address later, these three can be used together to dissect different roles of relaxation dynamics in TF–DNA assembly. For example, the comparison of the $\text{ETS}_1$ with the QSSA can reveal the effect of the time delay from the relaxation dynamics, while the result of the $\text{ETS}_2$ further reflects the effect of the amplitude reduction.

## III. Biochemical Applications

In Sec. II, we provided the revised versions of the MM rate law with the role of relaxation dynamics in complex formation, particularly relevant to time-varying molecular concentrations. We will apply these versions to the analysis of specific biomolecular systems.

### A. Protein–protein interaction

The tQSSA, rather than the sQSSA, has previously been recommended for the modeling of protein–protein interactions because the interacting proteins often show the concentrations comparable with each other [20,31]. Here, we will focus on the interactions between proteins whose concentrations oscillate over time with circadian rhythmicity, i.e., ~24-h periodicity. These time-varying concentrations can challenge the relevance of the tQSSA and require the examination by the ETS.

Consider the scenario that proteins A and B have oscillating concentrations in sinusoidal forms:

$$\overline{A}(\tau) = \overline{A}_{\max}\left\{1 - \frac{\alpha_{\mathrm{A}}}{2}\left[1 + \cos\left(\frac{2\pi}{k_\delta T}\tau\right)\right]\right\} \text{ and } \overline{B}(\tau) = \overline{B}_{\max}\left\{1 - \frac{\alpha_{\mathrm{B}}}{2}\left[1 + \cos\left(\frac{2\pi}{k_\delta T}\tau - \varphi_{\mathrm{B}}\right)\right]\right\}, \tag{34}$$

where $\overline{A}(\tau)$ and $\overline{B}(\tau)$ are the dimensionless A and B concentrations in Eq. (2), $\overline{A}_{\max}$ ($\overline{B}_{\max}$), $\alpha_{\mathrm{A(B)}}$, $T$, and $\varphi_{\mathrm{B}}$ denote the peak level of $\overline{A}(\tau)$ [$\overline{B}(\tau)$], the peak-to-trough difference of $\overline{A}(\tau)$ [$\overline{B}(\tau)$] divided by the peak level, the oscillation period of a circadian or diurnal rhythm, and the phase difference between $\overline{A}(\tau)$ and $\overline{B}(\tau)$, respectively. Here, $\alpha_{\mathrm{A}}$ and $\alpha_{\mathrm{B}}$ range from 0 to 1: the closer they are to 1, the stronger the oscillations. $\varphi_{\mathrm{B}}$ ranges from 0 to π without loss of generality. We here choose $T = 24$ h. Based on $\overline{A}(\tau)$ and $\overline{B}(\tau)$ in Eq. (34), we numerically solve Eq. (2) and thereby obtain $\bar{C}(\tau)$. We evaluate how well this $\bar{C}(\tau)$ is approximated by the $\mathrm{ETS}_1$, $\mathrm{ETS}_2$, and $\mathrm{ETS}_3$ in Eqs. (18), (19), and (21), the tQSSA in Eq. (5), and the sQSSA in Eq. (8).

As illustrated in Fig. 1(a), we observe that the $\mathrm{ETS}_1$, $\mathrm{ETS}_2$, and $\mathrm{ETS}_3$ tend to better match the temporal trajectory of $\bar{C}(\tau)$ than the tQSSA and sQSSA, but with some variations between them. To clarify this point, we define $\phi^t_{\mathrm{ETS}_1}$, $\phi^t_{\mathrm{ETS}_2}$, $\phi^t_{\mathrm{ETS}_3}$, $\phi^t_{\mathrm{tQ}}$, and $\phi^t_{\mathrm{sQ}}$ as the phase differences in hours between the $\mathrm{ETS}_1$ and $\bar{C}(\tau)$, between the $\mathrm{ETS}_2$ and $\bar{C}(\tau)$, between the $\mathrm{ETS}_3$ and $\bar{C}(\tau)$, between the tQSSA and $\bar{C}(\tau)$, and between the sQSSA and $\bar{C}(\tau)$, respectively (Appendix B). The sign of a given phase difference is assigned positive (negative) if the corresponding trajectory has a more advanced (delayed) phase than $\bar{C}(\tau)$. For example, the signs of $\phi^t_{\mathrm{tQ}}$ and $\phi^t_{\mathrm{sQ}}$ are mainly positive according to our observation. In the particular case in Fig. 1(a), $\phi^t_{\mathrm{ETS}_1} = -1.1$ h, $\phi^t_{\mathrm{ETS}_2} = 0.4$ h, $\phi^t_{\mathrm{ETS}_3} = -0.05$ h, $\phi^t_{\mathrm{tQ}} = 3.1$ h, and $\phi^t_{\mathrm{sQ}} = 3.1$ h. Hence, $|\phi^t_{\mathrm{ETS}_1}|$, $|\phi^t_{\mathrm{ETS}_2}|$, and $|\phi^t_{\mathrm{ETS}_3}|$ are smaller than $|\phi^t_{\mathrm{tQ}}|$ and $|\phi^t_{\mathrm{sQ}}|$.

---

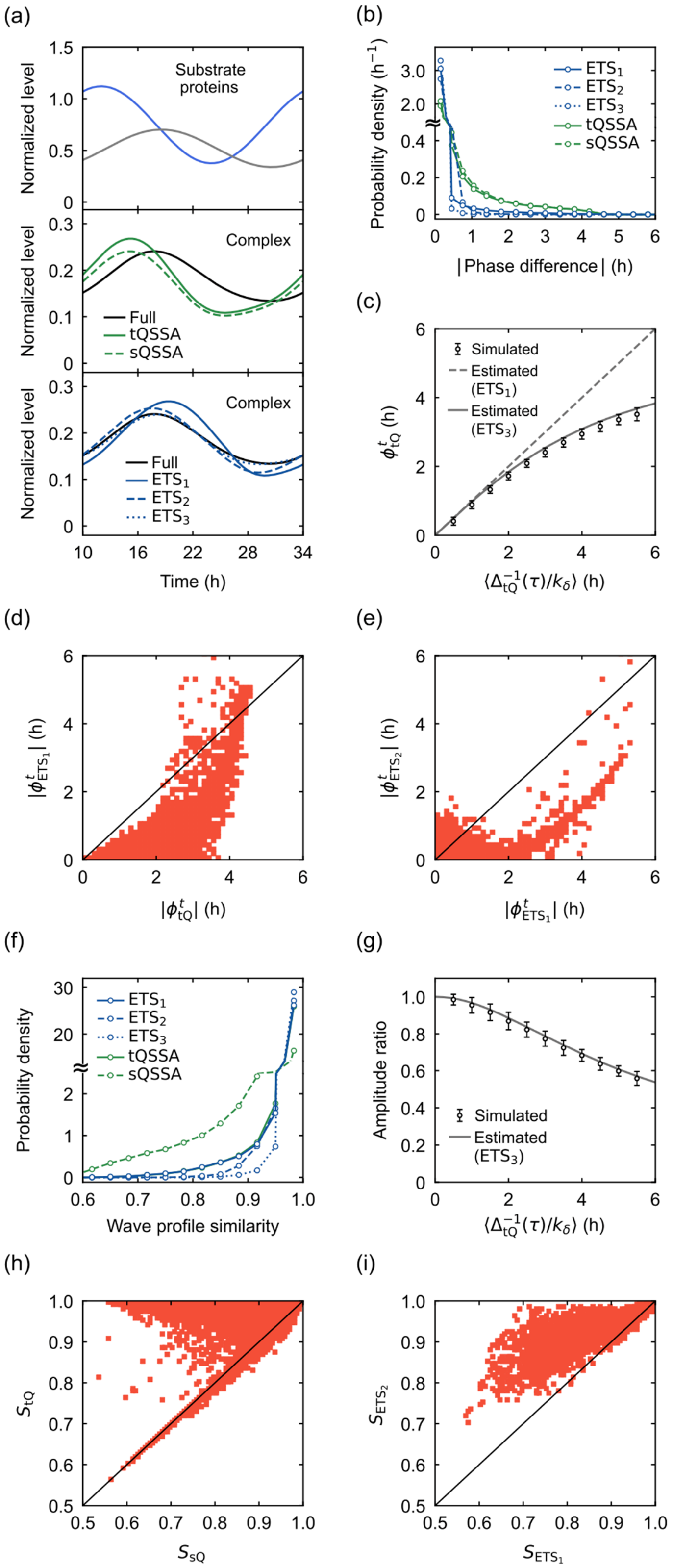

**Fig. 1. Protein–protein interaction.** (a) Example time-series of substrate protein levels $\bar{A}(\tau)$ and $\bar{B}(\tau)$ in Eq. (34) at the top, the full model-based complex level $\bar{C}(\tau)$ and the tQSSA and sQSSA at the center, and $\bar{C}(\tau)$ and the $ETS_1$, $ETS_2$, and $ETS_3$ at the bottom. $t = k_\delta^{-1}\tau$ as defined before Eq. (2). (b) Probability distributions of $|\phi_{\mathrm{ETS}_1}^t|$ ("$ETS_1$"), $|\phi_{\mathrm{ETS}_2}^t|$ ("$ETS_2$"), $|\phi_{\mathrm{ETS}_3}^t|$ ("$ETS_3$"), $|\phi_{\mathrm{tQ}}^t|$ ("tQSSA"), and $|\phi_{\mathrm{sQ}}^t|$ ("sQSSA") over randomly-sampled parameter sets. (c) Simulated versus analytically estimated $\phi_{\mathrm{tQ}}^t$ as a function of time-averaged $\Delta_{\mathrm{tQ}}^{-1}(\tau)/k_\delta$. The analytical estimate is the time delay in Eq. (18) ("$ETS_1$") or Eq. (21) ("$ETS_3$") with the time-averaged $\Delta_{\mathrm{tQ}}^{-1}(\tau)$. Each data point and error bar represent the average and standard deviation of the simulation results with randomly-sampled parameter sets. (d,e) Scatter plot of $|\phi_{\mathrm{tQ}}^t|$ and $|\phi_{\mathrm{ETS}_1}^t|$ (d), or that of $|\phi_{\mathrm{ETS}_1}^t|$ and $|\phi_{\mathrm{ETS}_2}^t|$ (e) with randomly-sampled parameter sets. A solid diagonal line corresponds to $|\phi_{\mathrm{ETS}_1}^t| = |\phi_{\mathrm{tQ}}^t|$ (d) or $|\phi_{\mathrm{ETS}_2}^t| = |\phi_{\mathrm{ETS}_1}^t|$ (e). (f) Probability distributions of $S_{\mathrm{ETS}_1}$ ("$ETS_1$"), $S_{\mathrm{ETS}_2}$ ("$ETS_2$"), $S_{\mathrm{ETS}_3}$ ("$ETS_3$"), $S_{\mathrm{tQ}}$ ("tQSSA"), and $S_{\mathrm{sQ}}$ ("sQSSA") over randomly-sampled parameter sets. (g) Simulated versus analytically estimated ratio of the full model-based complex level's amplitude to the tQSSA's, as a function of time-averaged $\Delta_{\mathrm{tQ}}^{-1}(\tau)/k_\delta$. The analytical estimate is the inverse of the square root term in Eq. (21) ("$ETS_3$") with the time-averaged $\Delta_{\mathrm{tQ}}^{-1}(\tau)$. Each data point and error bar represent the average and standard deviation of the simulation results with randomly-sampled parameter sets. (h,i) Scatter plot of $S_{\mathrm{sQ}}$ and $S_{\mathrm{tQ}}$ (h), or that of $S_{\mathrm{ETS}_1}$ and $S_{\mathrm{ETS}_2}$ (i) with randomly-sampled parameter sets. A solid diagonal line corresponds to $S_{\mathrm{tQ}} = S_{\mathrm{sQ}}$ (h) or $S_{\mathrm{ETS}_2} = S_{\mathrm{ETS}_1}$ (i). For parameters in (a) and (b)–(i), refer to Tables S4 and S1, respectively. For more details, refer to Sec. III.A and Appendix B.

---

We did find that $|\phi_{\mathrm{ETS}_1}^t|$, $|\phi_{\mathrm{ETS}_2}^t|$, and $|\phi_{\mathrm{ETS}_3}^t|$ tend to be smaller than $|\phi_{\mathrm{tQ}}^t|$ and $|\phi_{\mathrm{sQ}}^t|$ across physiologically-relevant conditions [Fig. 1(b), Table S1, and $P < 10^{-4}$]. As anticipated from the solid analytical basis of Eq. (21), the $ETS_3$ offers the closest phase to that of $\bar{C}(\tau)$, with $|\phi_{\mathrm{ETS}_3}^t| \le 1.6$ h across all our simulated conditions. This agreement is due to the correct functional form of the time delay in the $ETS_3$, which matches the actual phase difference between $\bar{C}(\tau)$ and its quasi-steady state ($\phi_{\mathrm{tQ}}^t$) even for long relaxation times, where the $ETS_1$'s delay does not [Fig. 1(c)]. We will then skip the further analysis of the $ETS_3$-predicted phase. We will also skip the use of $\phi_{\mathrm{sQ}}^t$ due to its high similarity with $\phi_{\mathrm{tQ}}^t$ (Spearman's $\rho = 0.92$ and $P < 10^{-4}$). Notably, when $|\phi_{\mathrm{ETS}_1}^t|$ or $|\phi_{\mathrm{tQ}}^t|$ is ≥1 h, most parameter conditions (99.0%) have $|\phi_{\mathrm{ETS}_1}^t|$ less than $|\phi_{\mathrm{tQ}}^t|$ [Fig. 1(d) and $P < 10^{-4}$]. There still exist parameter conditions with $|\phi_{\mathrm{ETS}_1}^t|$ ≥2 h and even ≥4 h, but such conditions are >13 times rarer for $|\phi_{\mathrm{ETS}_2}^t|$ [Fig. 1(e) and $P < 10^{-4}$]. In other words, both the $ETS_1$ and $ETS_2$ with the effect of the relaxation time provide a better estimate of the phase of $\bar{C}(\tau)$ than the quasi-steady state assumption, but the $ETS_2$ exhibits fewer noticeable errors than the $ETS_1$.

The latter comes from the fact that the $ETS_2$ predicts a rather smaller phase difference between $\bar{C}(\tau)$ and its quasi-steady state than the $ETS_1$, as the $ETS_3$ did in Fig. 1(c).

These findings demonstrate the importance of relaxation time in complex formation that should be accurately considered to deepen the understanding of the deviation of the complex profile from its quasi-steady state.

Other than phases, wave profiles themselves determined by the waveforms and peak levels are the important features of oscillations. Therefore, we calculate similarity $S_{\mathrm{ETS}_1}$ between the wave profiles of the $ETS_1$ and $\bar{C}(\tau)$ by aligning their phases to the same. $S_{\mathrm{ETS}_1}$ is devised to approach 1 away from 0, as the two wave profiles quantitatively better match each other (Appendix B). We also calculate the similarities $S_{\mathrm{ETS}_2}$, $S_{\mathrm{ETS}_3}$, $S_{\mathrm{tQ}}$, and $S_{\mathrm{sQ}}$ for the $ETS_2$ and $\bar{C}(\tau)$, for the $ETS_3$ and $\bar{C}(\tau)$, for the tQSSA and $\bar{C}(\tau)$, and for the sQSSA and $\bar{C}(\tau)$, respectively. Across physiologically-relevant conditions, we found that $S_{\mathrm{ETS}_2}$ and $S_{\mathrm{ETS}_3}$ tend to be larger than $S_{\mathrm{ETS}_1}$ and $S_{\mathrm{tQ}}$, but the latter ones still tend to be larger than $S_{\mathrm{sQ}}$ [Fig. 1(f), Table S1, and $P < 10^{-4}$]. As discussed in Sec. II.A, the amplitudes of complex profiles from the $ETS_2$ and $ETS_3$ undergo the reduction by relaxation processes, and this effect may account for the tendency of larger $S_{\mathrm{ETS}_2}$ and $S_{\mathrm{ETS}_3}$. The $ETS_3$ does provide the functional form of the reduced amplitude, which is in remarkable agreement with our simulation results [Fig. 1(g)]. On the other hand, as expected from Eq. (18), $S_{\mathrm{ETS}_1}$ and $S_{\mathrm{tQ}}$ are almost identical to each other (Spearman's $\rho = 0.89$ and $P < 10^{-4}$), and tend to exceed $S_{\mathrm{sQ}}$ [Fig. 1(h) and $P < 10^{-4}$] in a manner consistent with the previous suggestions that the tQSSA is more accurate than the sQSSA [20,31,35]. Consequently, $S_{\mathrm{ETS}_2}$ and $S_{\mathrm{ETS}_3}$ for each simulated condition tend to exceed its $S_{\mathrm{ETS}_1}$ and $S_{\mathrm{tQ}}$ [Fig. 1(i) and $P < 10^{-4}$]. Therefore, the amplitude reduction by the relaxation dynamics is a key feature of the wave profile of molecular complex with time-varying components.

**B. TF–DNA interaction**

We here consider TF–DNA interactions with circadian rhythmicity. Consider the scenario that the TF concentration oscillates over time with a sinusoidal form:

$$\overline{A}_{\mathrm{TF}}(\tau) = \overline{A}_{\max}\left\{1 - \frac{\alpha_{\mathrm{A}}}{2}\left[1 + \cos\left(\frac{2\pi}{k_{\mathrm{TF}\delta}T}\tau\right)\right]\right\}, \tag{35}$$

where $\overline{A}_{\mathrm{TF}}(\tau)$, $\overline{A}_{\max}$, $\alpha_{\mathrm{A}}$, and $T$ are the dimensionless TF concentration in Eq. (24), the peak level of $\overline{A}_{\mathrm{TF}}(\tau)$, the peak-to-trough difference of $\overline{A}_{\mathrm{TF}}(\tau)$ divided by the peak level, and the oscillation period of a circadian or diurnal rhythm, respectively. Here, $\alpha_{\mathrm{A}}$ ranges from 0 to 1 (the closer it is to 1, the stronger the oscillation) and we choose $T = 24$ h. Based on $\overline{A}_{\mathrm{TF}}(\tau)$ in Eq. (35), we numerically solve Eq. (24) and thereby obtain $\overline{C}_{\mathrm{TF}}(\tau)$. We evaluate how well this $\overline{C}_{\mathrm{TF}}(\tau)$ is approximated by the ETS$_1$, ETS$_2$, and ETS$_3$ in Eqs. (31)–(33) or by the QSSA in Eq. (25).

---

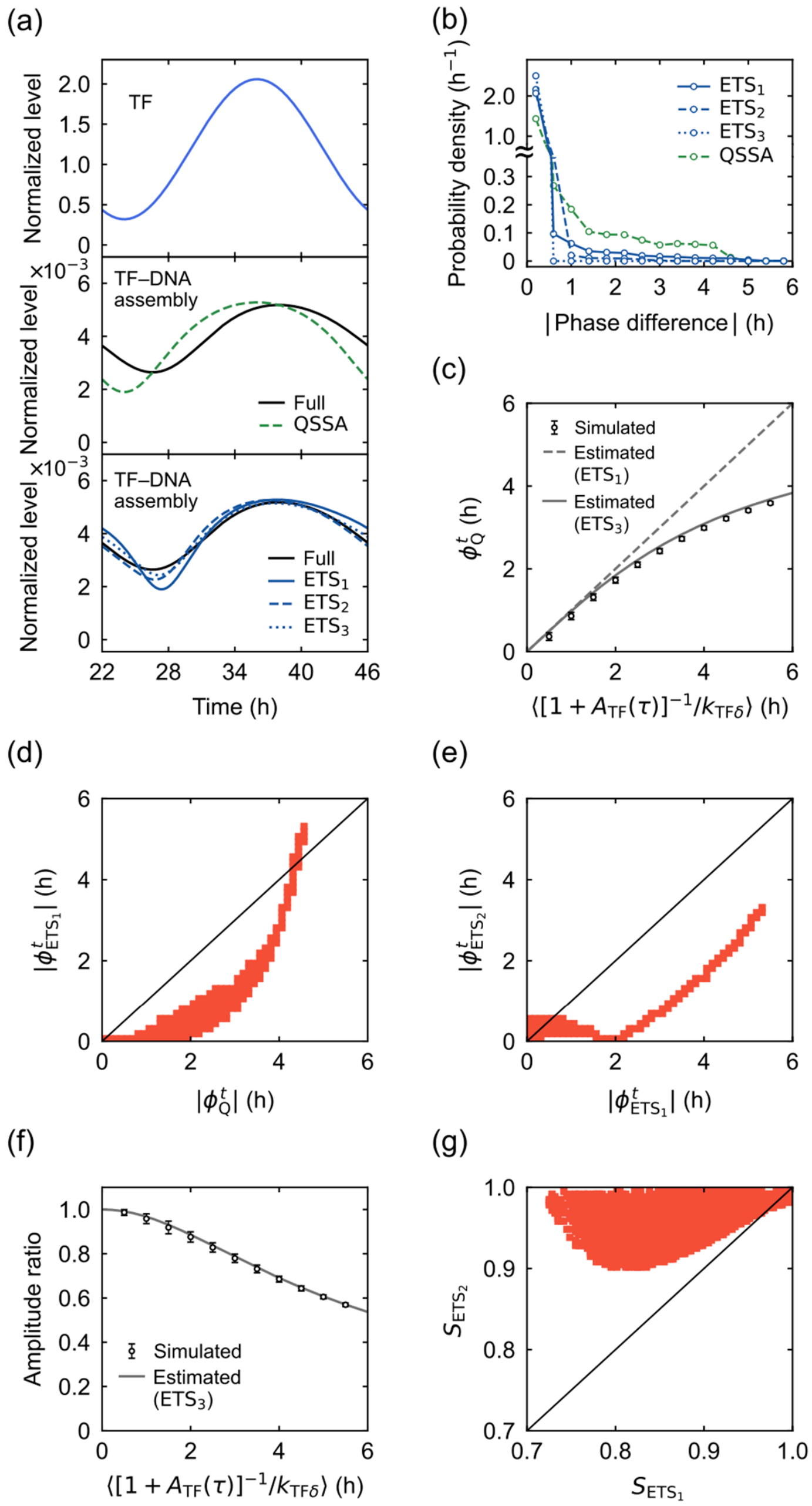

**Fig. 2. TF–DNA interaction.** (a) Example time-series of TF level $\overline{A}_{\mathrm{TF}}(\tau)$ in Eq. (35) at the top, the full model-based TF–DNA assembly level $\overline{C}_{\mathrm{TF}}(\tau)$ and the QSSA at the center, and $\overline{C}_{\mathrm{TF}}(\tau)$ and the ETS$_1$, ETS$_2$, and ETS$_3$ at the bottom. $t = k_\delta^{-1}\tau$ as defined under Eq. (24). (b) Probability distributions of $\left|\phi^t_{\mathrm{ETS}_1}\right|$ ("ETS$_1$"), $\left|\phi^t_{\mathrm{ETS}_2}\right|$ ("ETS$_2$"), $\left|\phi^t_{\mathrm{ETS}_3}\right|$ ("ETS$_3$"), and $\left|\phi^t_{\mathrm{Q}}\right|$ ("QSSA") over randomly-sampled parameter sets. (c) Simulated versus analytically estimated $\phi^t_{\mathrm{Q}}$ as a function of time-averaged

$[1+\bar{A}_{\mathrm{TF}}(\tau)]^{-1}/k_{\mathrm{TF\delta}}$. The analytical estimate is the time delay in Eq. (31) ("$\mathrm{ETS_1}$") or Eq. (33) ("$\mathrm{ETS_3}$") with the time-averaged $[1+\bar{A}_{\mathrm{TF}}(\tau)]^{-1}$. Each data point and error bar represent the average and standard deviation of the simulation results with randomly-sampled parameter sets. (d,e) Scatter plot of $|\phi_{\mathrm{Q}}^{t}|$ and $|\phi_{\mathrm{ETS_1}}^{t}|$ (d), or $|\phi_{\mathrm{ETS_1}}^{t}|$ and $|\phi_{\mathrm{ETS_2}}^{t}|$ (e) with randomly-sampled parameter sets. A solid diagonal line corresponds to $|\phi_{\mathrm{ETS_1}}^{t}| = |\phi_{\mathrm{Q}}^{t}|$ (d) or $|\phi_{\mathrm{ETS_2}}^{t}| = |\phi_{\mathrm{ETS_1}}^{t}|$ (e). (f) Simulated versus analytically estimated ratio of the full model-based TF–DNA assembly level's amplitude to the QSSA's, as a function of time-averaged $[1+\bar{A}_{\mathrm{TF}}(\tau)]^{-1}/k_{\mathrm{TF\delta}}$. The analytical estimate is the inverse of the square root term in Eq. (33) ("$\mathrm{ETS_3}$") with the time-averaged $[1+\bar{A}_{\mathrm{TF}}(\tau)]^{-1}$. Each data point and error bar represent the average and standard deviation of the simulation results with randomly-sampled parameter sets. (g) Scatter plot of $S_{\mathrm{ETS_1}}$ and $S_{\mathrm{ETS_2}}$ with randomly-sampled parameter sets. A solid diagonal line corresponds to $S_{\mathrm{ETS_2}} = S_{\mathrm{ETS_1}}$. For parameters in (a) and (b)–(g), refer to Tables S5 and S1, respectively. For more details, refer to Sec. III.B and Appendix B.

---

As illustrated in Fig. 2(a), we observe that the $\mathrm{ETS_1}$, $\mathrm{ETS_2}$, and $\mathrm{ETS_3}$ tend to better match the time trajectory of $\bar{C}_{\mathrm{TF}}(\tau)$ than the QSSA. For systematic evaluation, we define $\phi_{\mathrm{ETS_1}}^{t}$, $\phi_{\mathrm{ETS_2}}^{t}$, $\phi_{\mathrm{ETS_3}}^{t}$, and $\phi_{\mathrm{Q}}^{t}$ as the phase differences in hours between the $\mathrm{ETS_1}$ and $\bar{C}_{\mathrm{TF}}(\tau)$, between the $\mathrm{ETS_2}$ and $\bar{C}_{\mathrm{TF}}(\tau)$, between the $\mathrm{ETS_3}$ and $\bar{C}_{\mathrm{TF}}(\tau)$, and between the QSSA and $\bar{C}_{\mathrm{TF}}(\tau)$, respectively (Appendix B). The sign of a given phase difference is assigned positive (negative) if the corresponding trajectory has a more advanced (delayed) phase than $\bar{C}_{\mathrm{TF}}(\tau)$. It is observed that $\phi_{\mathrm{Q}}^{t} \geq 0$ across the simulated conditions (Table S1). In the example of Fig. 2(a), $\phi_{\mathrm{ETS_1}}^{t} = -0.6$ h, $\phi_{\mathrm{ETS_2}}^{t} = 0.4$ h, $\phi_{\mathrm{ETS_3}}^{t} = -0.1$ h, and $\phi_{\mathrm{Q}}^{t} = 2.3$ h. Hence, $|\phi_{\mathrm{ETS_1}}^{t}|$, $|\phi_{\mathrm{ETS_2}}^{t}|$, and $|\phi_{\mathrm{ETS_3}}^{t}|$ are smaller than $|\phi_{\mathrm{Q}}^{t}|$. Indeed, our analysis suggests that $|\phi_{\mathrm{ETS_1}}^{t}|$, $|\phi_{\mathrm{ETS_2}}^{t}|$, and $|\phi_{\mathrm{ETS_3}}^{t}|$ tend to be smaller than $|\phi_{\mathrm{Q}}^{t}|$ over physiologically-relevant conditions [Fig. 2(b) and $P < 10^{-4}$]. As expected from the solid analytical basis of Eq. (33), the $\mathrm{ETS_3}$ offers the closest phase to that of $\bar{C}_{\mathrm{TF}}(\tau)$ with $|\phi_{\mathrm{ETS_3}}^{t}| \leq 0.35$ h across all the simulated conditions. This agreement is due to the correct functional form of the time delay in the $\mathrm{ETS_3}$, which matches the actual phase difference between $\bar{C}_{\mathrm{TF}}(\tau)$ and its quasi-steady state ($\phi_{\mathrm{Q}}^{t}$) even for long relaxation times, where the $\mathrm{ETS_1}$'s delay does not [Fig. 2(c)]. We will then skip the further analysis of the $\mathrm{ETS_3}$-predicted phase. When $|\phi_{\mathrm{ETS_1}}^{t}|$ or $|\phi_{\mathrm{Q}}^{t}|$ is ≥1 h, most parameter conditions (97.7%) have $|\phi_{\mathrm{ETS_1}}^{t}|$ less than $|\phi_{\mathrm{Q}}^{t}|$ [Fig. 2(d) and $P < 10^{-4}$]. Notably, we found that $|\phi_{\mathrm{ETS_2}}^{t}|$ is rarely ≥ 1 h, 4.5 times rarer than $|\phi_{\mathrm{ETS_1}}^{t}| \geq 1$ h [Fig. 2(e) and $P < 10^{-4}$]. In other words, both the $\mathrm{ETS_1}$ and $\mathrm{ETS_2}$ with the effect of relaxation time account for the phase of $\bar{C}_{\mathrm{TF}}(\tau)$ deviating from the quasi-steady state assumption, but the $\mathrm{ETS_2}$ better does than the $\mathrm{ETS_1}$: the $\mathrm{ETS_2}$ predicts a rather smaller phase difference between $\bar{C}_{\mathrm{TF}}(\tau)$ and its quasi-steady state than the $\mathrm{ETS_1}$, as the $\mathrm{ETS_3}$ did in Fig. 2(c). This result suggests that the accurate consideration of the relaxation time in TF–DNA binding deepens the understanding of the deviation of $\bar{C}_{\mathrm{TF}}(\tau)$ from the quasi-steady state.

Apart from phases, we can examine the wave profiles predicted by the $\mathrm{ETS}_1$ by calculating similarity $S_{\mathrm{ETS}_1}$ between the profiles of the $\mathrm{ETS}_1$ and $\bar{C}_{\mathrm{TF}}(\tau)$ by aligning their phases to the same. $S_{\mathrm{ETS}_1}$ is devised to approach 1 away from 0, as the two wave profiles quantitatively better match each other (Appendix B). We also calculate the similarities $S_{\mathrm{ETS}_2}$, $S_{\mathrm{ETS}_3}$, and $S_{\mathrm{Q}}$ for the $\mathrm{ETS}_2$ and $\bar{C}_{\mathrm{TF}}(\tau)$, for the $\mathrm{ETS}_3$ and $\bar{C}_{\mathrm{TF}}(\tau)$, and for the QSSA and $\bar{C}_{\mathrm{TF}}(\tau)$, respectively. As anticipated from Eq. (31), $S_{\mathrm{ETS}_1}$ and $S_{\mathrm{Q}}$ take almost the same values as each other (Spearman's $\rho = 0.94$ and $P < 10^{-4}$). On the other hand, as discussed in Sec. II.B, the $\mathrm{ETS}_2$ and $\mathrm{ETS}_3$ reflect the amplitude reduction by relaxation processes and thus may give more accurate wave profiles than the $\mathrm{ETS}_1$ and QSSA. The $\mathrm{ETS}_3$ does provide the functional form of the reduced amplitude, which is in remarkable agreement with our simulation results [Fig. 2(f)]. Also, we found that both $S_{\mathrm{ETS}_2}$ and $S_{\mathrm{ETS}_3}$ exceed 0.9 for all our simulated conditions, whereas $S_{\mathrm{ETS}_1}$ and $S_{\mathrm{Q}}$ do not [Fig. 2(g) and $P < 10^{-4}$]. Therefore, the amplitude reduction by the relaxation dynamics is a key feature of the TF–DNA assembly profile with time-varying TF concentration.

### C. Mammalian circadian clock

The quantitatively more accurate results from the ETS than from the quasi-steady state assumption prompt us to ask whether the features in the ETS can also account for dynamical patterns that are qualitatively distinct from their quasi-steady states. We will answer this question through the analysis of the mammalian circadian system.

The core part of the mammalian circadian clock harbors a transcriptional/post-translational negative feedback loop that generates autonomous protein oscillations with circadian rhythmicity [32,40]. Heterodimers of CLOCK and BMAL1 proteins activate the transcription of *Period* (*Per*) and *Cryptochrome* (*Cry*) genes, and the encoded PER and CRY proteins form PER–CRY complexes that are translocated to the nucleus. In the nucleus, they interact with CLOCK–BMAL1 complexes to inhibit the CLOCK–BMAL1 transcriptional activities. These positive (CLOCK and BMAL1) and negative (PER and CRY) arms constitute the negative feedback loop for circadian oscillations in protein levels and activities.

A previous study suggests that the tQSSA for the interaction between activator (CLOCK–BMAL1) and repressor (PER–CRY) leads to more natural rhythm generation than the sQSSA, because the tQSSA captures the ultrasensitive response of the repressor's transcription to the activator or repressor's concentration—that is, a large change in the transcription rate against a small change in the activator or repressor level [31]. This ultrasensitive response, which is manifested by small $K$ and balance of activator and repressor levels, amplifies the rhythms and prevents their dampening [30,31]. Here, we will show that the $ETS_1$ and $ETS_2$ further capture the intrinsic time delay and amplitude reduction in the protein–protein and protein–DNA assembly formation and thereby reveal rhythmic patterns unexpected by the tQSSA. In this analysis, we will not use the $ETS_3$ because of its limitation, requiring prior knowledge of the oscillation period itself.

For the modeling of the mammalian clock, we interpret $A(t)$, $B(t)$, and $C(t)$ in Eq. (1) as the concentrations of activator, repressor, and their complex in the nucleus, respectively. For simplicity, we assume the constancy of $A(t)$, i.e., $A(t) = A$ as the activator's oscillation is relatively weak and dispensable for the circadian rhythmicity [40–43]. The resulting model comprises Eq. (1) with $A(t) = A$ and the following equations modified from a previous model [31]:

$$\frac{\mathrm{d}M(t)}{\mathrm{d}t} = a_0 C_{\mathrm{TF}}(t) - b_0 M(t), \tag{36}$$

$$\frac{\mathrm{d}B_{\mathrm{cy}}(t)}{\mathrm{d}t} = a_1 M(t) - b_1 B_{\mathrm{cy}}(t), \tag{37}$$

$$\frac{\mathrm{d}B(t)}{\mathrm{d}t} = a_2 B_{\mathrm{cy}}(t) - r_{\mathrm{f}}[B(t) - C(t)] - r_{\mathrm{c}} C(t), \tag{38}$$

$$\frac{\mathrm{d}C_{\mathrm{TF}}(t)}{\mathrm{d}t} = \frac{k_{\mathrm{TFa}}}{V}[A - C(t)] - \{k_{\mathrm{TF\delta}} + k_{\mathrm{TFa}}[A - C(t)]\} C_{\mathrm{TF}}(t). \tag{39}$$

Here, $M(t)$, $B_{\mathrm{cy}}(t)$, and $C_{\mathrm{TF}}(t)$ are the concentrations of repressor mRNA, cytoplasmic repressor protein, and activator on repressor's promoter, respectively. Eq. (39) is equivalent to Eq. (24) in its content. $a_0$, $a_1$, and $a_2$ denote the transcription, translation, and translocation rates of the repressor, respectively. $b_1$ represents the sum of the translocation and degradation rates of the repressor in the cytoplasm, and thus satisfies $b_1 N_{\mathrm{c}} > a_2$ where $N_{\mathrm{c}}$ is the cytoplasm-to-nucleus volume ratio. $b_0$, $r_{\mathrm{f}}$, and $r_{\mathrm{c}}$ are the degradation rates of repressor mRNA, free repressor protein, and activator-binding repressor protein, respectively. By definition, $r_{\mathrm{c}}$ satisfies $r_{\mathrm{c}} < k_{\delta}$ for $k_{\delta}$ in Eq. (1). $k_{\mathrm{TFa}}$, $k_{\mathrm{TF\delta}}$, and $V$ are those in Eqs. (22)–(24). To be precise, $A - C(t)$ in Eq. (1) should be replaced by $A - C(t) - C_{\mathrm{TF}}(t)$; however, this replacement does not much affect our simulation results, and thus we keep the original form of Eq. (1) for the straightforward use of the approximants for $C(t)$ as will be shown later.

Our model simulation with Eqs. (1) and (36)–(39) leads to the oscillation of the variables in a subset of the parameter conditions in Table S3 (Appendix B). For comparison, we test other versions of the model with the $ETS_1$ and $ETS_2$. The model with the $ETS_1$ ($ETS_2$) only consists of Eqs. (36)–(38) where $C(t)$ and $C_{\mathrm{TF}}(t)$ are respectively replaced by $K\bar{C}_{\gamma_1}(t)$ $[K\bar{C}_{\gamma_2}(t)]$ and $K_{\mathrm{TF}}\bar{C}_{\mathrm{TF}\gamma_1}(t)$ $[K_{\mathrm{TF}}\bar{C}_{\mathrm{TF}\gamma_2}(t)]$ from Eqs. (18) and (31) [Eqs. (19) and (32)], and $A - K\bar{C}_{\gamma_1}(t)$ $[A - K\bar{C}_{\gamma_2}(t)]$ corresponds to $A_{\mathrm{TF}}(t)$ in Eqs. (25) and (31). Likewise, the tQSSA- and sQSSA-based models are respectively constructed by using the approximants in Eqs. (5) and (8) instead of $\bar{C}_{\gamma_1}(t)$ or $\bar{C}_{\gamma_2}(t)$, and using $\bar{C}_{\mathrm{TFQ}}(t)$ in Eq. (25) instead of $\bar{C}_{\mathrm{TF}\gamma_1}(t)$ or $\bar{C}_{\mathrm{TF}\gamma_2}(t)$.

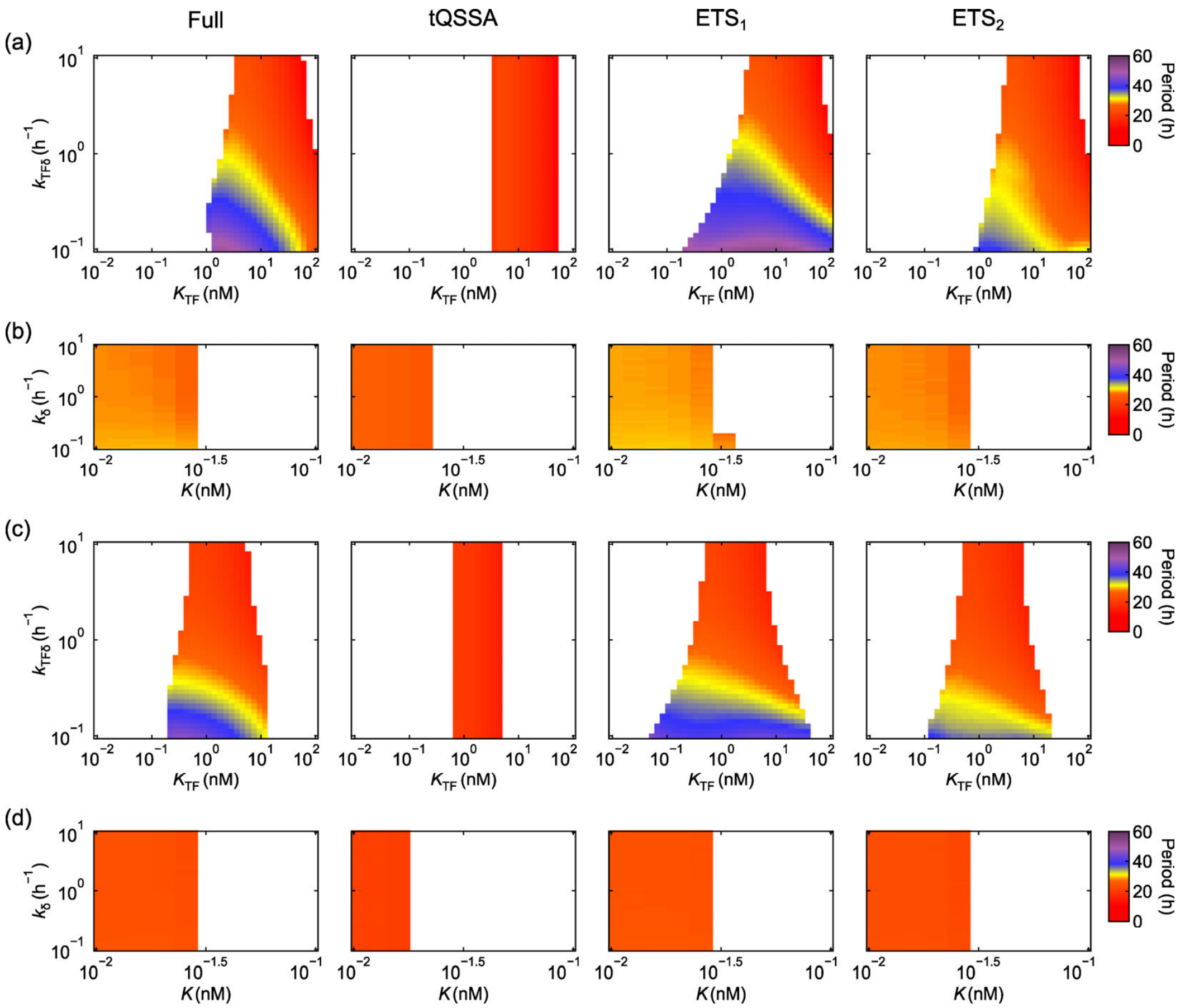

**Fig. 3. Mammalian circadian clock.** In (a)–(d), the parameter ranges with observed oscillations from each simulation scheme are colored by their oscillation periods, according to the color scale on the rightmost side. (a) and (b) have the same parameters except $K_{\mathrm{TF}}$, $k_{\mathrm{TF}\delta}$, $K$, and $k_{\delta}$. Likewise, (c) and (d) have the same parameters except $K_{\mathrm{TF}}$, $k_{\mathrm{TF}\delta}$, $K$, and $k_{\delta}$. The simulation results from the full model with Eqs. (1) and (36)–(39) (left), from the tQSSA (center left), from the $\mathrm{ETS}_1$ (center right), and from the $\mathrm{ETS}_2$ (right) are presented over the ranges of $K_{\mathrm{TF}}$ and $k_{\mathrm{TF}\delta}$ in (a) and (c), or the ranges of $K$ and $k_{\delta}$ in (b) and (d), while the other parameters are fixed. For more details, refer to Sec. III.C, Appendix B, and Table S6.

Consistent with the previous work and the characteristics of the well-studied Goodwin model [31,44], the sQSSA-based model fails to produce oscillations for any simulated conditions (Appendix B and Table S3). On the other hand, the tQSSA-based model generates oscillations as previously demonstrated [31], but with substantial deviations from the exact model simulation with Eqs. (1) and (36)–(39) (Fig. 3 and Appendix B). For example, a decrease in $k_{\mathrm{TF}\delta}$ in the exact model simulation tends to facilitate the development of oscillations to some extent by widening the oscillatory range of $K_{\mathrm{TF}}$, and concurrently lengthens the oscillation period [Figs. 3(a) and 3(c)]; the tQSSA-based model does not reproduce these patterns [Figs. 3(a) and 3(c)]. Besides, the exact simulation results exhibit the oscillations over the range of $K$ beyond small $K$ values from the tQSSA-based results [Figs. 3(b) and 3(d)]. Moreover, the tQSSA-based oscillation periods are limited to short periods compared with the exact simulation results (Fig. 3).

In contrast, the simulation with the $\mathrm{ETS}_1$ shows reasonable agreement with the exact model simulation for the overall patterns of oscillation onset and periods (Fig. 3 and Appendix B). For example, the $\mathrm{ETS}_1$ predicts the periods of 23.9 to 52.3 h with varying $k_{\mathrm{TF}\delta}$ at $K_{\mathrm{TF}} = 3.7$ nM in Fig. 3(a), while the exact periods span 22.8 to 48.6 h in the same conditions. Meanwhile, the tQSSA-based period there is limited to 19.5 h [Fig. 3(a)]. Regarding the $\mathrm{ETS}_1$, the wide range of the oscillatory parameters and the period variations comparable with the exact simulation results comes from the time delay in protein–protein and protein–DNA assembly formation. This intrinsic time-delay effect, which is absent in the tQSSA, enhances the rhythmicity and lengthens the oscillation period. In this respect, the $\mathrm{ETS}_2$-based model produces similar patterns to the $\mathrm{ETS}_1$'s as well (Fig. 3).

A caveat of the $\mathrm{ETS}_1$ is that it tends to overestimate the oscillatory parameter range compared to the exact model simulation [Figs. 3(a) and 3(c)]. This error is notably corrected by the simulation of the $\mathrm{ETS}_2$-based model, whose amplitude reduction effect tends to dampen the oscillation and thereby narrow the oscillatory parameter range closer to the exact range. In the example of Fig. 3(c), the $\mathrm{ETS}_1$ predicts an 18.1%-wider range of oscillatory parameters than the exact range, whereas the $\mathrm{ETS}_2$ exhibits only a 3.3% difference from that exact range. About the $\mathrm{ETS}_2$, this smaller oscillatory regime is also partially attributed to its shorter, overall time delay than the $\mathrm{ETS}_1$'s, as apparent from Eqs. (18), (19), (31), and (32); however, this short time delay alone does not fully reduce the oscillatory regime to the level of the $\mathrm{ETS}_2$ with the amplitude reduction effect (Fig. 4 and Appendix C). Yet, the $\mathrm{ETS}_2$ is not void of a downside: it sometimes generates little yet spurious humps in a molecular time-series apart from its genuine peaks (Fig. 5), because of the superposition of two oscillatory curves in Eqs. (19) and (32).

Taken together, the joint effect of the time delay and amplitude reduction from the relaxation dynamics determines the emergence of oscillations as shown by the $\mathrm{ETS}_2$, and this time delay further sets the period variations of the oscillations as shown by the $\mathrm{ETS}_1$ and $ETS_2$. Therefore, both the time delay and amplitude reduction act crucial to the qualitative and quantitative aspects of the circadian model dynamics, unexpected by the quasi-steady state assumption.

---

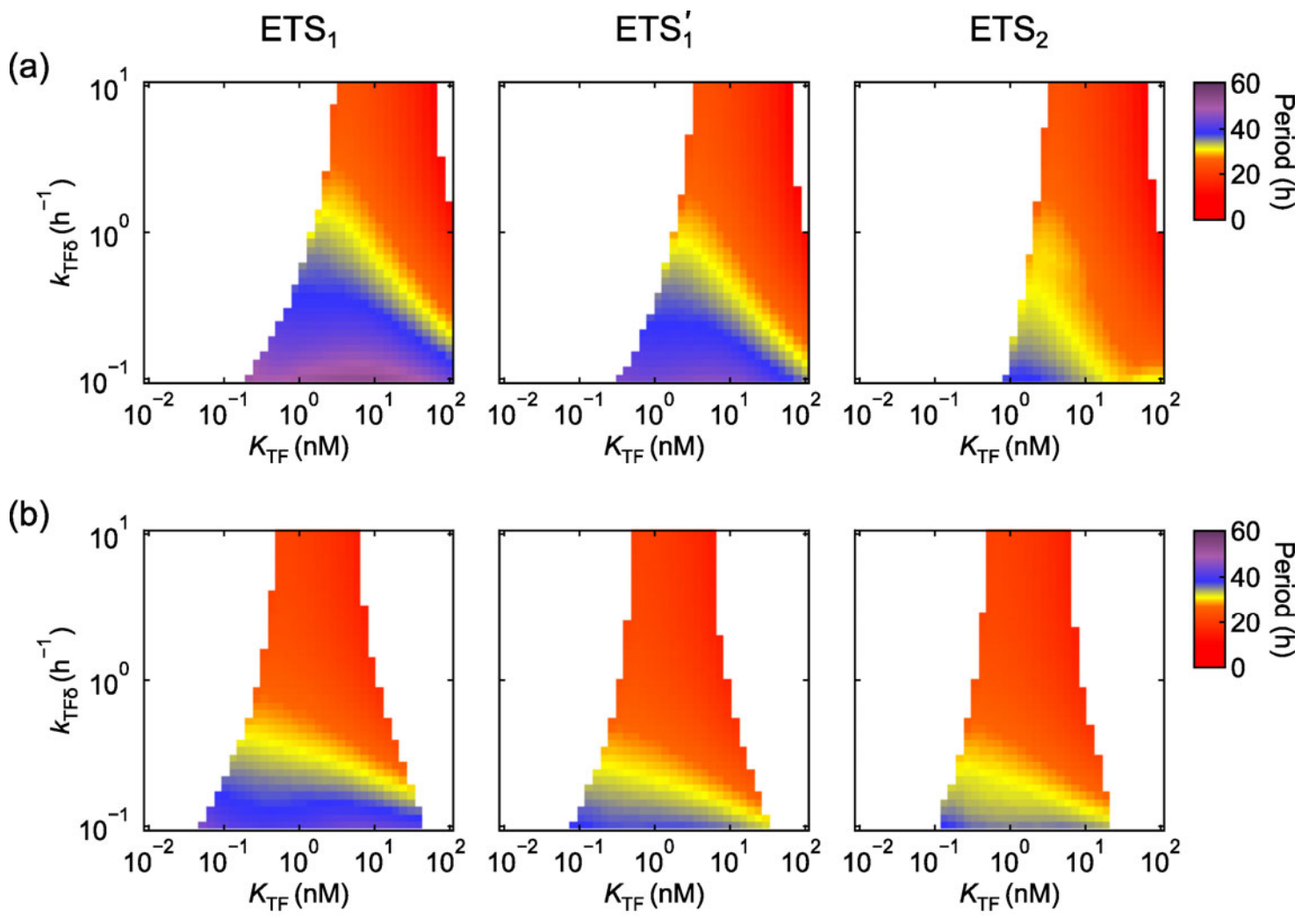

**Fig. 4. Shortening the time delay without the amplitude reduction effect for the mammalian circadian clock.** (a) and (b) involve the same parameters as Figs. 3(a) and 3(c), respectively. Presented are the results from the $ETS_1$ (left) and $ETS_2$ (right), in addition to those from $\bar{C}_{\gamma_1'}(\tau)$ and $\bar{C}_{\mathrm{TF}\gamma_1'}(\tau)$ in Eqs. (C1) and (C2) (center, "$ETS_1'$") designed for the time delay similar to the $ETS_2$ but without the amplitude reduction. For more details, refer to Sec. III.C and Appendices B and C.

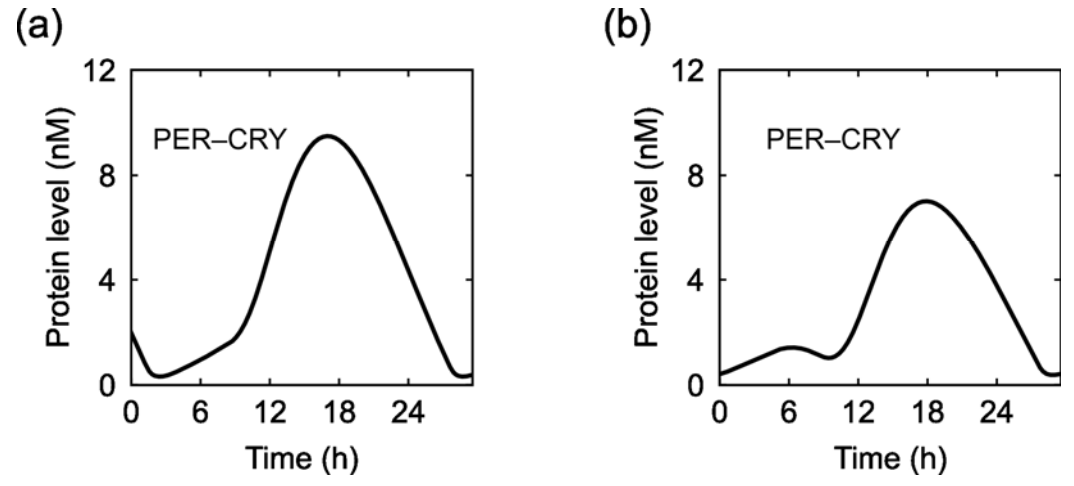

**Fig. 5. The $ETS_2$-based time-series of the mammalian clock component, nuclear and cytoplasmic PER–CRY complex concentration.** Typically, the $ETS_2$-result shows a single peak in each period (a), but sometimes an additional hump (b) that is absent in the full-model simulation with Eqs. (1) and (36)–(39). In (a) and (b), $K_{\mathrm{TF}} = 38.88$ nM, $k_{\mathrm{TF}\delta} = 0.16$ $h^{-1}$ (a) or $0.10$ $h^{-1}$ (b), and the other parameters are the same as Fig. 3(a).

## IV. Conclusion and Discussion

The deviation of time-varying molecular activities from their quasi-steady states is possibly prevalent but often overlooked with the use of the MM rate law or its modified forms [10–22,37,38]. We previously investigated this issue and suggested that the relaxation time of molecular complex formation, quantitated by the effective time delay, contributes to the breakdown of the quasi-steady state assumption [35]. This time delay aside, our current study reveals that the amplitude reduction from the relaxation dynamics is another important factor for the deviation of a molecular trajectory from the quasi-steady state. The time delay and amplitude reduction do not only affect the phase and wave profile of the complex but also impact the qualitative output of the system, such as the onset of oscillations and their period variations in the mammalian circadian clock model. Hence, unless the quasi-steady state assumption is reasonably held like the case in Eq. (15) or (29), the relaxation dynamics should be considered for both the qualitative and quantitative descriptions of molecular interaction systems with time-varying components. The relaxation dynamics can be considered by the simulation of complex-complete mass-action models (full kinetic models without any approximants for complex concentrations) or by the ETS for analytical insights as demonstrated in Figs. 1(c), 1(g), 2(c), and 2(f).

Incorporating the effect of stochastic fluctuations in molecular binding events [4,36–39,45,46] into the future analysis, beyond the TF–DNA interactions examined in this study, would enhance the understanding of molecular interaction dynamics with time-varying components. This part may be achieved by extending the currently-available stochastic QSSA [37,38]. Further simulation and experimental profiling of molecular complexes in various regulatory or signaling pathways [23,24,35,47] would help establish the role of the relaxation dynamics raised in this study. Experimental data with high temporal resolution would allow their precise comparison with the ETS-based complex profiles. In such endeavors, protein complexes can be quantified by densitometric analysis with western blotting, enzyme-linked immunosorbent assay, mass spectrometry-based proteomics with co-immunoprecipitation, and other related techniques.

## Appendix A: Amplitude Reduction by the $\mathrm{ETS}_2$

About the $\mathrm{ETS}_2$ with Eq. (19) [Eq. (32)], we will prove its amplitude reduction compared to the tQSSA (QSSA). For the $\mathrm{ETS}_2$ with Eq. (19), we start with the following relations:

$$\begin{aligned}\min_\tau\left[\bar{C}_{\mathrm{tQ}}(\tau)\right] &= e^{-1}\min_\tau\left[\bar{C}_{\mathrm{tQ}}(\tau)\right] + (1-e^{-1})\min_\tau\left\{\bar{C}_{\mathrm{tQ}}\left[\tau - \Delta_{\mathrm{tQ}}^{-1}(\tau)\right]\right\} \\ &\le e^{-1}\bar{C}_{\mathrm{tQ}}(\tau) + (1-e^{-1})\bar{C}_{\mathrm{tQ}}\left[\tau - \Delta_{\mathrm{tQ}}^{-1}(\tau)\right],\end{aligned} \tag{A1}$$

$$\min_\tau\left[\bar{C}_{\mathrm{tQ}}(\tau)\right] \le \bar{C}_{\mathrm{tQ}}(\tau) \le \min[\bar{A}(\tau), \bar{B}(\tau)]. \tag{A2}$$

The relation $\bar{C}_{\mathrm{tQ}}(\tau) \le \min[\bar{A}(\tau), \bar{B}(\tau)]$ in Eq. (A2) was found in Sec. II.A. From Eqs. (19), (A1), and (A2), we obtain $\min_\tau\left[\bar{C}_{\mathrm{tQ}}(\tau)\right] \le \bar{C}_{\gamma_2}(\tau)$ and then

$$\min_\tau\left[\bar{C}_{\mathrm{tQ}}(\tau)\right] \le \min_\tau\left[\bar{C}_{\gamma_2}(\tau)\right]. \tag{A3}$$

From $\bar{C}_{\gamma_2}(\tau) \le e^{-1}\bar{C}_{\mathrm{tQ}}(\tau) + (1-e^{-1})\bar{C}_{\mathrm{tQ}}\left[\tau - \Delta_{\mathrm{tQ}}^{-1}(\tau)\right]$, $\bar{C}_{\mathrm{tQ}}(\tau) \le \max_\tau\left[\bar{C}_{\mathrm{tQ}}(\tau)\right]$, and $\bar{C}_{\mathrm{tQ}}\left[\tau - \Delta_{\mathrm{tQ}}^{-1}(\tau)\right] \le \max_\tau\left[\bar{C}_{\mathrm{tQ}}(\tau)\right]$, we obtain

$$\max_\tau\left[\bar{C}_{\gamma_2}(\tau)\right] \le \max_\tau\left[\bar{C}_{\mathrm{tQ}}(\tau)\right]. \tag{A4}$$

As evident from Eqs. (A3) and (A4), the amplitude of $\bar{C}_{\gamma_2}(\tau)$ can never exceed that of $\bar{C}_{\mathrm{tQ}}(\tau)$.

Next, in the case of the $\mathrm{ETS}_2$ with Eq. (32), $\min_\tau\left[\bar{C}_{\mathrm{TFQ}}(\tau)\right] \le \bar{C}_{\mathrm{TFQ}}(\tau)$ and $\min_\tau\left[\bar{C}_{\mathrm{TFQ}}(\tau)\right] \le \bar{C}_{\mathrm{TFQ}}\{\tau - [1 + \bar{A}_{\mathrm{TF}}(\tau)]^{-1}\}$ lead to

$$\min_\tau\left[\bar{C}_{\mathrm{TFQ}}(\tau)\right] \le \min_\tau\left[\bar{C}_{\mathrm{TF}\gamma_2}(\tau)\right]. \tag{A5}$$

Similarly, $\bar{C}_{\mathrm{TFQ}}(\tau) \le \max_\tau\left[\bar{C}_{\mathrm{TFQ}}(\tau)\right]$ and $\bar{C}_{\mathrm{TFQ}}\{\tau - [1 + \bar{A}_{\mathrm{TF}}(\tau)]^{-1}\} \le \max_\tau\left[\bar{C}_{\mathrm{TFQ}}(\tau)\right]$ lead to

$$\max_\tau\left[\bar{C}_{\mathrm{TF}\gamma_2}(\tau)\right] \le \max_\tau\left[\bar{C}_{\mathrm{TFQ}}(\tau)\right]. \tag{A6}$$

As evident from Eqs. (A5) and (A6), the amplitude of $\bar{C}_{\mathrm{TF}\gamma_2}(\tau)$ can never exceed that of $\bar{C}_{\mathrm{TFQ}}(\tau)$.

## Appendix B: Numerical Methods

Numerical simulations and analyses were performed by Python 3.7.0 or 3.7.4. Ordinary differential equations (ODEs) were solved by LSODA (scipy.integrate.solve_ivp) in SciPy v1.1.0 or v1.3.1 with the maximum time step of 0.05 h. Delay differential equations were solved by a modified version of the ddeint module with LSODA [48].

For the parameter selection in numerical simulations or for the null model generation in statistical significance tests, random numbers were sampled by the Mersenne Twister in random.py.

Spearman's $\rho$ was measured with scipy.stats.spearmanr in SciPy v1.1.0 or v1.3.1. To test the significance of $\rho$ between two groups of variables, we randomized the pairing of the variables between the groups (while maintaining the original group membership) and measured the $P$ value (one-tailed) from $10^4$ null configurations.

In Secs. III.A and III.B, $10^5$ parameter sets were randomly selected for Eqs. (2) and (34) or for Eqs. (24) and (35), according to Table S1. We found that the simulated $\bar{C}(\tau)$ and $\bar{C}_{\mathrm{TF}}(\tau)$ were insensitive to their initial conditions. A phase difference between two periodic time-series was calculated by maximizing their cross-correlation with a varying displacement of one series relative to the other [49]. For the cross-correlation calculation, the time average of each series was shifted to zero and ten duplicates of a single time period ($10 \times T$) was used [35]. The cross-correlation was obtained with scipy.signal.correlate in SciPy v1.1.0 or v1.3.1 (mode = 'same' and method = 'fft'). To test the significance of the observation in Sec. III.A that $|\phi_{\mathrm{ETS}_1}^t|$ tends to be smaller than $|\phi_{\mathrm{tQ}}^t|$, we obtained $\Lambda = \mathrm{med}(|\phi_{\mathrm{tQ}}^t|) - \mathrm{med}(|\phi_{\mathrm{ETS}_1}^t|)$ [where $\mathrm{med}(\cdots)$ is the median over the parameter sets with $|\phi_{\mathrm{tQ}}^t| \geq 1$ h or $|\phi_{\mathrm{ETS}_1}^t| \geq 1$ h] and randomized the labeling of $\bar{C}_{\gamma_1}(\tau)$ and $\bar{C}_{\mathrm{tQ}}(\tau)$ for each parameter set. We then measured $\Lambda_{\mathrm{r}} = \mathrm{med}(|\phi_{\mathrm{tQ}}^t|) - \mathrm{med}(|\phi_{\mathrm{ETS}_1}^t|)$ with the new sets of $|\phi_{\mathrm{tQ}}^t|$ and $|\phi_{\mathrm{ETS}_1}^t|$ in this null configuration. The $P$ value was given by the probability of $\Lambda_{\mathrm{r}} \geq \Lambda$ across the $10^4$ null configurations. The analogous methods were applied to the other similar comparisons in Secs. III.A and III.B. In Sec. III.A, we tested the significance of the fraction ($q$) of parameter sets with $|\phi_{\mathrm{ETS}_1}^t| < |\phi_{\mathrm{tQ}}^t|$ as follows, when $|\phi_{\mathrm{ETS}_1}^t|$ or $|\phi_{\mathrm{tQ}}^t|$ is ≥1 h: for each parameter set, we randomized the labeling of $\bar{C}_{\gamma_1}(\tau)$ and $\bar{C}_{\mathrm{tQ}}(\tau)$, and measured the fraction ($q_{\mathrm{r}}$) of the parameter sets of $|\phi_{\mathrm{ETS}_1}^t| < |\phi_{\mathrm{tQ}}^t|$ with the new $|\phi_{\mathrm{ETS}_1}^t|$ and $|\phi_{\mathrm{tQ}}^t|$ when $|\phi_{\mathrm{ETS}_1}^t|$ or $|\phi_{\mathrm{tQ}}^t|$ is ≥1 h. The $P$ value was given by the probability of $q_{\mathrm{r}} \geq q$ across the $10^4$ null configurations. The analogous method was applied to the other similar case in Sec. III.B. Likewise, in Sec. III.A, we tested the significance of the fraction ($q'$) of parameter sets with $|\phi_{\mathrm{ETS}_2}^t| \geq v$ ($v =$ 2h or 4h) compared to the case of $|\phi_{\mathrm{ETS}_1}^t|$, as follows: for each parameter set, we randomized the labeling of $\bar{C}_{\gamma_2}(\tau)$ and $\bar{C}_{\gamma_1}(\tau)$, and measured the fraction ($q'_{\mathrm{r}}$) of the parameter sets of $|\phi^t_{\mathrm{ETS}_2}| \geq v$ with the new $|\phi^t_{\mathrm{ETS}_2}|$. The $P$ value was given by the probability of $q' \geq q'_{\mathrm{r}}$ across the $10^4$ null configurations. The analogous method was applied to the other similar case in Sec. III.B.

We measured the similarity between two time-series $f_1(t)$ and $f_2(t)$ as $\left\{\int_t^{t+T} \min[f_1(t'), f_2(t')]\mathrm{d}t'\right\} / \int_t^{t+T} \max[f_1(t'), f_2(t')]\mathrm{d}t'$, where $T$ is an oscillation period of $f_1(t)$ and $f_2(t)$ [33]. This quantity takes a range of 0 to 1, and becomes large for quantitatively similar profiles of $f_1(t)$ and $f_2(t)$. In the case of wave profile similarities in Secs. III.A and III.B, the phase difference between the two time-series in the comparison was set to zero by the phase shift of the one series to the other, before measuring their wave profile similarity. To test the significance of the observation in Sec. III.A that $S_{\mathrm{ETS}_2}$ tends to be larger than $S_{\mathrm{ETS}_1}$, we followed a similar procedure to the above phase-difference comparison, but used the parameter sets with $S_{\mathrm{ETS}_2} \leq 0.9$ or $S_{\mathrm{ETS}_1} \leq 0.9$. This method was also applied to the other related comparisons in Sec. III.A. We further tested the significance of the fraction of parameter sets with $S_{\mathrm{ETS}_2} > S_{\mathrm{ETS}_1}$ in a similar way to the above phase-difference analysis, but only based on the parameter sets with $S_{\mathrm{ETS}_2} \leq 0.9$ or $S_{\mathrm{ETS}_1} \leq 0.9$. This method was also applied to the other related analyses in Sec. III.A. In Sec. III.B, we tested the significance of the fraction of parameter sets with $S_{\mathrm{ETS}_2} > 0.9$ compared to the case of $S_{\mathrm{ETS}_1}$ in a similar way to the above phase-difference analysis, and this method was also applied to the other related analyses.

In Sec. III.C, $10^5$ parameter sets from Table S3 were randomly selected for simulation. The simulation results were insensitive to the initial conditions of variables. The oscillation of the system in its asymptotic behavior was heuristically defined to meet all the following conditions at the end of the simulation: (*i*) $\alpha \geq 0.2$ where $\alpha$ is the peak-to-trough difference divided by the peak level (the peak and trough levels here are the averages of three consecutive peak and trough levels, respectively), (*ii*) differences in three consecutive peak levels are smaller than 5% of the highest peak level, and (*iii*) differences in three consecutive periods are smaller than 5% of the longest period. Here, the period is measured by the average of three consecutive peak-time differences. Peaks and sub-peaks (humps) were identified with scipy.signal.find_peaks in SciPy v1.3.1. Although the above conditions (*i*)–(*iii*) cannot capture non-periodic oscillations, essentially all the observed oscillations in Sec. III.C were periodic. Each oscillation was monitored by the concentration profile of PER–CRY complex in the nucleus and cytoplasm.

## Appendix C: Shortening the Effective Time Delay

To examine the pure effect of a shorter time delay with the $ETS_2$ than with the $ETS_1$, one should first generate a similar time delay to $\bar{C}_{\gamma_2}(\tau)$ in Eq. (19) but without the amplitude reduction effect. In this respect, we try the following approximant $\bar{C}_{\gamma_1'}(\tau)$ for $\bar{C}_{\gamma_2}(\tau)$:

$$\bar{C}_{\gamma_1'}(\tau) \equiv \min\{\bar{C}_{\mathrm{tQ}}[\tau - (1 - e^{-1})\Delta_{\mathrm{tQ}}^{-1}(\tau)], \bar{A}(\tau), \bar{B}(\tau)\}. \quad \text{(C1)}$$

$\bar{C}_{\gamma_1'}(\tau)$ is almost free of the amplitude reduction and reminiscent of $\bar{C}_{\gamma_1}(\tau)$ [Eq. (18)] in its form. The reason for the use of $\bar{C}_{\gamma_1'}(\tau)$ is that $\bar{C}_{\gamma_2}(\tau) \approx \bar{C}_{\mathrm{tQ}}(\tau) - (1 - e^{-1})\Delta_{\mathrm{tQ}}^{-1}(\tau)\bar{C}'(\tau) \approx \bar{C}_{\gamma_1'}(\tau)$ for small $\Delta_{\mathrm{tQ}}^{-1}(\tau)$, from the Taylor expansion of $\bar{C}_{\mathrm{tQ}}(\tau - x)$ by $x = \Delta_{\mathrm{tQ}}^{-1}(\tau)$ or $(1 - e^{-1})\Delta_{\mathrm{tQ}}^{-1}(\tau)$. The phase difference in hours between $\bar{C}_{\mathrm{tQ}}(\tau)$ and $\bar{C}_{\gamma_2}(\tau)$ does well match the time average of $(1 - e^{-1})\Delta_{\mathrm{tQ}}^{-1}(\tau)/k_\delta$ from $\bar{C}_{\gamma_1'}(\tau)$ (Spearman's $\rho = 0.995$ and $P < 10^{-4}$), according to the model simulation in Sec. III.A.

In the case of the time delay close to $\bar{C}_{\mathrm{TF}\gamma_2}(\tau)$ in Eq. (32), a similar mathematical reason to the above is also applied to the following approximant $\bar{C}_{\mathrm{TF}\gamma_1'}(\tau)$ for $\bar{C}_{\mathrm{TF}\gamma_2}(\tau)$:

$$\bar{C}_{\mathrm{TF}\gamma_1'}(\tau) \equiv \bar{C}_{\mathrm{TFQ}}\left[\tau - \frac{1-e^{-1}}{1+\bar{A}_{\mathrm{TF}}(\tau)}\right]. \quad \text{(C2)}$$

The phase difference in hours between $\bar{C}_{\mathrm{TFQ}}(\tau)$ and $\bar{C}_{\mathrm{TF}\gamma_2}(\tau)$ does well match the time average of $(1 - e^{-1})[1 + \bar{A}_{\mathrm{TF}}(\tau)]^{-1}/k_{\mathrm{TF}\delta}$ from $\bar{C}_{\mathrm{TF}\gamma_1'}(\tau)$ (Spearman's $\rho = 0.99$ and $P < 10^{-4}$), according to the model simulation in Sec. III.B.

## Acknowledgments

This work was supported by the National Research Foundation of Korea (RS-2023-00263411) and the UBSI Research Fund of Ulsan National Institute of Science and Technology (No. 1.230070.01) (J.C., R.L., and C.-M.G.). We also acknowledge the support of the Blue Sky Research Fund of Hong Kong Baptist University (RC-BSRF/21-22/09) and the General Research Fund from the Research Grants Council of the Hong Kong Special Administrative Region, China (No. 12202322) (R.L., T.L.P.M., and P.-J.K.). This work was conducted with the resources of the UNIST Supercomputing Center and the HKBU High Performance Cluster Computing Centre.

## References

1. M. Born and R. Oppenheimer, *Zur Quantentheorie der Molekeln*, Ann. Phys. **389**, 457–484 (1927).
2. J. Monod, J. Wyman, and J.-P. Changeux, *On the Nature of Allosteric Transitions: A Plausible Model*, J. Mol. Biol. **12**, 88–118 (1965).
3. G. K. Ackers, A. D. Johnson, and M. A. Shea, *Quantitative Model for Gene Regulation by λ Phage Repressor*, Proc. Natl. Acad. Sci. USA **79**, 1129–1133 (1982).
4. H.-W. Kang and T. G. Kurtz, *Separation of Time-Scales and Model Reduction for Stochastic Reaction Networks*, Ann. Appl. Probab. **23**, 529–583 (2013).
5. F. G. Heineken, H. M. Tsuchiya, and R. Aris, *On the Mathematical Status of the Pseudo-Steady State Hypothesis of Biochemical Kinetics*, Math. Biosci. **1**, 95–113 (1967).
6. L. A. Segel and M. Slemrod, *The Quasi-Steady-State Assumption: a Case Study in Perturbation*, SIAM Rev. **31**, 446–477 (1989).
7. J. W. Dingee and A. B. Anton, *A New Perturbation Solution to the Michaelis–Menten Problem*, AIChE J. **54**, 1344–1357 (2008).
8. A. M. Bersani, E. Bersani, G. Dell'Acqua, and M. G. Pedersen, *New Trends and Perspectives in Nonlinear Intracellular Dynamics: One Century from Michaelis–Menten Paper*, Continuum. Mech. Thermodyn. **27**, 659–684 (2015).
9. G. E. Briggs and J. B. S. Haldane, *A Note on the Kinetics of Enzyme Action*, Biochem. J. **19**, 338 (1925).
10. A. R. Tzafriri, *Michaelis–Menten Kinetics at High Enzyme Concentrations*, Bull. Math. Biol. **65**, 1111–1129 (2003).
11. V. Henri, *Lois Générales de l'Action des Diastases* (Librairie Scientifique A. Hermann, 1903).

12. L. Michaelis and M. L. Menten, *Die Kinetik der Invertinwirkung*, Biochem. Z. **49**, 333–369 (1913).
13. C. Gérard and A. Goldbeter, *A Skeleton Model for the Network of Cyclin-Dependent Kinases Driving the Mammalian Cell Cycle*, Interface Focus **1**, 24–35 (2011).
14. K. C. Chen, A. Csikasz-Nagy, B. Gyorffy, J. Val, B. Novak, and J. J. Tyson, *Kinetic Analysis of a Molecular Model of the Budding Yeast Cell Cycle*, Mol. Biol. Cell **11**, 369–391 (2000).
15. J. C. Leloup and A. Goldbeter, *Toward a Detailed Computational Model for the Mammalian Circadian Clock*, Proc. Natl. Acad. Sci. USA **100**, 7051–7056 (2003).
16. J. Garcia-Ojalvo, *Modeling Gene Expression in Time and Space*, Annu. Rev. Biophys. **42**, 605–627 (2013).
17. M. Foo, D. E. Somers, and P.-J. Kim, *Kernel Architecture of the Genetic Circuitry of the Arabidopsis Circadian System*, PLOS Comput. Biol. **12**, e1004748 (2016).
18. T. D. Pollard, *A Guide to Simple and Informative Binding Assays*, Mol. Biol. Cell **21**, 4061–4067 (2010).
19. A. D. Attie and R. T. Raines, *Analysis of Receptor–Ligand Interactions*, J. Chem. Educ. **72**, 119 (1995).
20. J. K. Kim and J. J. Tyson, *Misuse of the Michaelis–Menten Rate Law for Protein Interaction Networks and Its Remedy*, PLOS Comput. Biol. **16**, e1008258 (2020).
21. S. Schnell, *Validity of the Michaelis–Menten Equation – Steady-State or Reactant Stationary Assumption: That is the Question*, FEBS J. **281**, 464–472 (2014).
22. J. Eilertsen, S. Schnell, and S. Walcher, *On the Anti-Quasi-Steady-State Conditions of Enzyme Kinetics*, Math. Biosci. **350**, 108870 (2022).
23. A. Fujioka, K. Terai, R. E. Itoh, K. Aoki, T. Nakamura, S. Kuroda *et al.*, *Dynamics of the Ras/ERK MAPK Cascade as Monitored by Fluorescent Probes*, J. Biol. Chem. **281**, 8917–8926 (2006).
24. N. Blüthgen, F. J. Bruggeman, S. Legewie, H. Herzel, H. V. Westerhoff, and B. N. Kholodenko, *Effects of Sequestration on Signal Transduction Cascades*, FEBS J. **273**, 895–906 (2006).
25. S. Carmi, E. Y. Levanon, S. Havlin, and E. Eisenberg, *Connectivity and Expression in Protein Networks: Proteins in a Complex Are Uniformly Expressed*, Phys. Rev. E. **73**, 031909 (2006).
26. K. J. Laidler, *Theory of the Transient Phase in Kinetics, with Special Reference to Enzyme Systems*, Can. J. Chem. **33**, 1614–1624 (1955).

27. S. Cha, *Kinetic Behavior at High Enzyme Concentrations: Magnitude of Errors of Michaelis–Menten and Other Approximations*, J. Biol. Chem. **245**, 4814–4818 (1970).
28. H. C. Lim, *On Kinetic Behavior at High Enzyme Concentrations*, AICHE J. **19**, 659–661 (1973).
29. J. A. M. Borghans, R. J. de Boer, and L. A. Segel, *Extending the Quasi-Steady State Approximation by Changing Variables*, Bull. Math. Biol. **58**, 43–63 (1996).
30. N. E. Buchler and M. Louis, *Molecular Titration and Ultrasensitivity in Regulatory Networks*, J. Mol. Biol. **384**, 1106–1119 (2008).
31. J. K. Kim and D. B. Forger, *A Mechanism for Robust Circadian Timekeeping via Stoichiometric Balance*, Mol. Syst. Biol. **8**, 630 (2012).
32. F. Gachon, E. Nagoshi, S. Brown, J. Ripperger, and U. Schibler, *The Mammalian Circadian Timing System: From Gene Expression to Physiology*, Chromosoma **113**, 103–112 (2004).
33. R. Lim, J. Chae, D. E. Somers, C.-M. Ghim, and P.-J. Kim, *Cost-Effective Circadian Mechanism: Rhythmic Degradation of Circadian Proteins Spontaneously Emerges without Rhythmic Post-Translational Regulation*, iScience **24**, 102726 (2021).
34. N. Mosheiff, B. M. C. Martins, S. Pearl-Mizrahi, A. Grünberger, S. Helfrich, I. Mihalcescu, D. Kohlheyer, J. C. W. Locke, L. Glass, and N. Q. Balaban, *Inheritance of Cell-Cycle Duration in the Presence of Periodic Forcing*, Phys. Rev. X **8**, 021035 (2018).
35. R. Lim, T. L. P. Martin, J. Chae, W. J. Kim, C.-M. Ghim, and P.-J. Kim, *Generalized Michaelis–Menten Rate Law with Time-Varying Molecular Concentrations*, PLOS Comput. Biol. **19**, e1011711 (2023).
36. V. Kampen and N. Godfried, *Stochastic Processes in Physics and Chemistry* (Elsevier, 1992).
37. J. K. Kim and E. D. Sontag, *Reduction of Multiscale Stochastic Biochemical Reaction Networks Using Exact Moment Derivation*, PLOS Comput. Biol. **13**, e1005571 (2017).
38. Y. M. Song, H. Hong, and J. K. Kim, *Universally Valid Reduction of Multiscale Stochastic Biochemical Systems Using Simple Non-Elementary Propensities*, PLOS Comput. Biol. **17**, e1008952 (2021).
39. E. Levine and T. Hwa, *Stochastic Fluctuations in Metabolic Pathways*, Proc. Natl. Acad. Sci. USA **104**, 9224–9229 (2007).
40. C. Lee, J. P. Etchegaray, F. R. Cagampang, A. S. Loudon, and S. M. Reppert, *Posttranslational Mechanisms Regulate the Mammalian Circadian Clock*, Cell **107**, 855–867 (2001).

41. A. C. Liu *et al*., *Redundant Function of REV-ERBα and β and Non-Essential Role for Bmal1 Cycling in Transcriptional Regulation of Intracellular Circadian Rhythms*, PLoS Genet. **4**, e1000023 (2008).
42. N. Preitner *et al*., *The Orphan Nuclear Receptor REV-ERBα Controls Circadian Transcription within the Positive Limb of the Mammalian Circadian Oscillator*, Cell **110**, 251–260 (2002).
43. E. L. McDearmon *et al*., *Dissecting the Functions of the Mammalian Clock Protein BMAL1 by Tissue-Specific Rescue in Mice*, Science **314**, 1304–1308 (2006).
44. B. C. Goodwin, *Oscillatory Behavior in Enzymatic Control Processes*, Adv. Enzyme Regul. **3**, 425–438 (1965).
45. C.-M. Ghim and E. Almaas, *Two-Component Genetic Switch as a Synthetic Module with Tunable Stability*, Phys. Rev. Lett. **103**, 028101 (2009).
46. P.-J. Kim and N. D. Price, *Macroscopic Kinetic Effect of Cell-to-Cell Variation in Biochemical Reactions*, Phys. Rev. Lett. **104**, 148103 (2010).
47. H.-H. Jo, Y. J. Kim, J. K. Kim, M. Foo, D. E. Somers, and P.-J. Kim, *Waveforms of Molecular Oscillations Reveal Circadian Timekeeping Mechanisms*, Commun. Biol. **1**, 207 (2018).
48. Scipy-based delay differential equation (dde) solver, https://github.com/Zulko/ddeint.
49. H. Wang and P. Chu, V*oice Source Localization for Automatic Camera Pointing System in Video Conferencing*, 1997 IEEE International Conference on Acoustics, Speech, and Signal Processing **1** , 187–190 (1997).